\documentclass[12pt, aps,prd,onecolumn,showpacs,floats,floatfix,letterpaper,nofootinbib,notitlepage]{revtex4-1}
\pdfoutput=1
\usepackage{hyperref}
\usepackage{amssymb,amsmath,latexsym,mathrsfs}
\usepackage{graphicx,subfigure}
\usepackage{epsfig}
\usepackage{varioref,xr-hyper}
\usepackage{color}
\usepackage{multirow}
\usepackage{array}
\usepackage{tcolorbox}

\hypersetup{
	colorlinks   = true,
	citecolor    = violet,
	linkcolor = cyan
}

\newcommand\Tstrut{\rule{0pt}{2.6ex}}

\begin{document}

\title{On the convergence of cosmographic expansions in Lemaitre-Tolman-Bondi models}

\author{Asha B. Modan}
\email{ashabmodan@cp3.sdu.dk}
\author{S. M. Koksbang}
\email{koksbang@cp3.sdu.dk}

\affiliation{$CP^3$-Origins, University of Southern Denmark, Campusvej 55, DK-5230 Odense M, Denmark}
%\ead{ashabmodan@cp3.sdu.dk, koksbang@cp3.sdu.dk}

\begin{abstract}
	We study cosmographic expansions of the luminosity distance for a variety of Lemaitre-Tolman-Bondi models which we specify inspired by local large-scale structures of the universe. We consider cosmographic expansions valid for general spacetimes and compare to the Friedmann-Lemaitre-Robertson-Walker (FLRW) limit of the expansions as well as to its naive isotropic extrapolation to an inhomogeneous universe. The FLRW expansions are often poor near the observer but become better at higher redshifts, where the light rays have reached the FLRW background. In line with this we find that the effective Hubble, deceleration and jerk parameters of the general cosmographic expansion are often very different from the global $\Lambda$CDM values, with deviations up to several orders of magnitude. By comparing with the naive isotropic extrapolation of the FLRW expansion, we assess that these large deviations are mainly due to gradients of the shear. Very close to the observer, the general cosmographic expansion is always best and becomes more precise when more expansion terms are included. However, we find that the convergence radius of the general cosmographic expansion is small for all studied models and observers and the general cosmographic expansion becomes poor for most of the studied observers already before a single LTB structure has been traversed. The small radius of convergence of the general cosmographic expansion has also been indicated by earlier work and may need careful attention before we can safely apply the general cosmographic expansion to real data.
\end{abstract}
\maketitle

\section{Introduction}
Nearly 100 years ago, observations making it clear that the Universe is expanding were presented \cite{Lemaitre, Slipher, Hubble}. Today, the cosmic expansion history as well as $H_0$ are important sources of information about the Universe. One of the main observables in cosmology is the redshift-distance relation which in the Friedmann-Lemaitre-Robertson-Walker (FLRW) models has a simple relation to the expansion rate through the integral $\int_0^zdz/H(z)$. By measuring luminosity distances to e.g. supernovae we can therefore learn about the cosmic expansion history. However, just as was done in its original detection, cosmic expansion is often considered through a Taylor series expansion of the redshift-distance relation. This type of Taylor series expansion is referred to as a cosmographic expansion and plays a central role in modern cosmology. Cosmographic expansions are for instance routinely used for low-redshift measurements of the Hubble constant $H_0$ (see e.g. \cite{H0_1, H0_2, H0_3, H0_4, H0_5, H0_6, H0_7}), and in general for studies using various observations related to cosmic expansion (see e.g. \cite{ex_1, ex_2, ex_3, ex_4, ex_5, ex_6, ex_7, ex_8} for some examples). Cosmographic expansions are also used specifically for constraining the equation-of-state parameter of dark energy and/or alternative theories of gravity \cite{eos_1, eos_2, eos_3, eos_4, eos_5, eos_6, Fr_gravity} (but see also \cite{critique_1} for shortcomings), and can be used for constraining more exotic scenarios such as models with a varying speed of light \cite{varying_light} or fine structure constant \cite{fine}.
\newline\indent
Originally, cosmographic expansions were performed as Taylor expansions in the redshift \cite{jerk_1, jerk_2} (see \cite{see_also} for similar early considerations), but expansions in other parameters have been explored to obtain more precise expansions, especially to increase the precision of cosmographic expansions at higher redshift (see e.g. \cite{better_1, better_2, better_3, better_4, better_5, better_6, better_7, better_8} for various considerations and comparisons). While the standard approach to cosmography is valid only for spatially homogeneous and isotropic spacetimes, studies have also considered expansions for more general spacetimes. Especially after the appearance of the Hubble tension \cite{tension_1, tension_2, tension_3}, efforts \cite{general_1, general_2, general_3} have been put into constructing cosmographic expansions based on inhomogeneous spacetimes (see also e.g. \cite{early_1, early_2, early_3, early_4} for earlier similar considerations). Recent studies \cite{recent_1, recent_2, recent_3} indicate that observations are reaching a precision where interpretations of actual observations with these more general cosmographic expansions can yield information about our local cosmic environment. Put another way, the local cosmic environment may affect the interpretation of observations at detectable level, including interpretations based on cosmography. However, the use of cosmographic expansions for general spacetimes is still at its infancy and at the same time, this type of expansion naturally leads to significantly more complex expressions than expansions valid only in the FLRW limit. This means that it can be difficult to identify which contributions to the expansion coefficients are important and under what circumstances. In addition, the convergence behavior of the expansions can become more subtle in the inhomogeneous case. We therefore suggest that the current studies using general cosmographic expansions with e.g. simulation data be supplemented by studies using exact solutions to Einstein's equations which can provide well-understood simplified spacetimes where the importance of inhomogeneities and anisotropies for the expansions can be studied in a more controlled manner. This will be important for achieving a better understanding of under what circumstances the general versus FLRW cosmographic expansions are best used with real data. Motivated by this, we here study the convergence of the cosmographic expansions in specific Lemaitre-Tolman-Bondi (LTB) \cite{Lemaitre_LTB, Tolman, Bondi} models. We consider both FLRW-based cosmographic expansions and cosmographic expansions valid for general spacetimes.
\newline\indent
In section \ref{sec:model_setup} below we introduce the LTB models used in our study. Section \ref{sec:light} is dedicated to presenting the light propagation formalism we use for computing the exact redshift-distance relations as well as the different cosmographic expansions we consider. We present our results in section \ref{sec:results} where we include a comparison to earlier work. In section \ref{sec:conclusion} we give a summary and provide concluding remarks.

\section{Model setup}\label{sec:model_setup}
In this section we give a brief introduction to the LTB models before providing details on the individual LTB models we will consider in the following sections.
\newline\newline
The LTB models are dust-solutions to Einstein's equations with spherically symmetric spatial hypersurfaces orthogonal to the dust fluid flow. A cosmological constant can be included. Using spherical coordinates, the line element of the LTB models can be written as
\begin{align}
ds^2 = -c^2 dt^2 + R(t,r)dr^2 + A^2(t,r)\left(d\theta^2 + \sin^2(\theta)d\phi^2 \right), 
\end{align}
where $R = \frac{A^2_{,r}(t,r)}{1-k(r)}$ (a subscripted comma followed by a coordinate denotes partial derivative with respect to that coordinate). The LTB models evolve according to
\begin{equation}\label{eq:dynamical}
\frac{1}{c^2}A_{,t}^2 = \frac{2M}{A} - k +\frac{1}{3c^2}\Lambda A^2
\end{equation}
and the dust density of the models is given by
\begin{equation}
\rho = \frac{2M_{,r}}{c^2\beta A^2A_{,r}  },\, \, \, \, \, \, \, \, \beta = 8\pi G/c^4.
\end{equation}
The LTB models have two free functions in $r$ and in addition the models have coordinate covariance in $r$. The latter means that we can specify a third function of $r$ (which simply corresponds to rescaling $r$).
\newline\indent
To specify our LTB models we first set the Big Bang time to zero so that $t$ is the age of the model universe. The rescaling of the radial coordinate is then used to set $A(t_i, r) = a(t_i)r$, where $t_i$ is the age of the Universe when the redshift is $z = 1100$ in a flat $\Lambda$CDM model with $H_0 = 70$km/s/Mpc and $\Omega_{m,0} = 0.3$. This $\Lambda$CDM model will be referred to as the background model. The final specification of the model is obtained by specifying $k(r)$ which we do in the subsection below, where we specify six LTB models. The different models specified below represent different versions of our local environment, where we vary e.g. the local density at various degrees, guided by observational evidence as detailed below.

\begin{figure}[]
	\centering
	\includegraphics[width=0.45\linewidth]{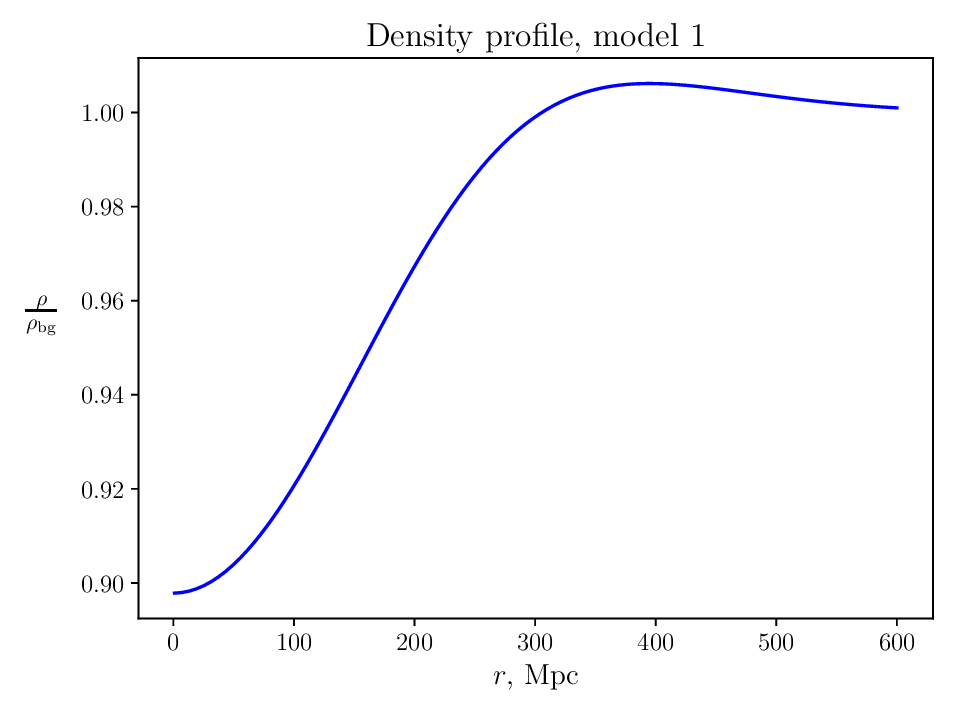}%
	\includegraphics[width=0.45\linewidth]{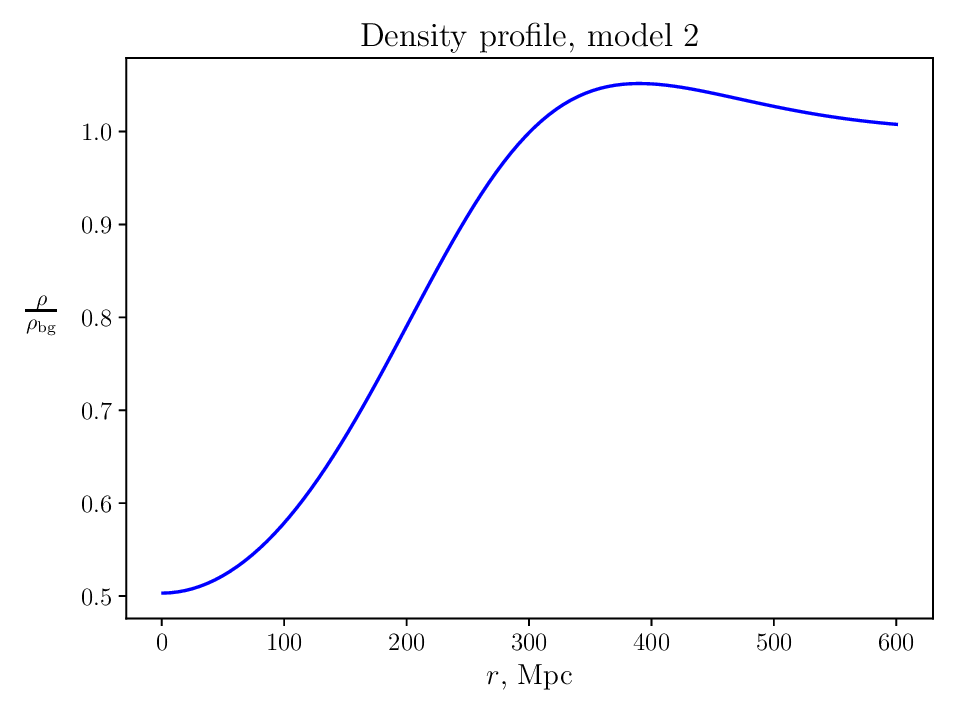}\\
	\includegraphics[width=0.45\linewidth]{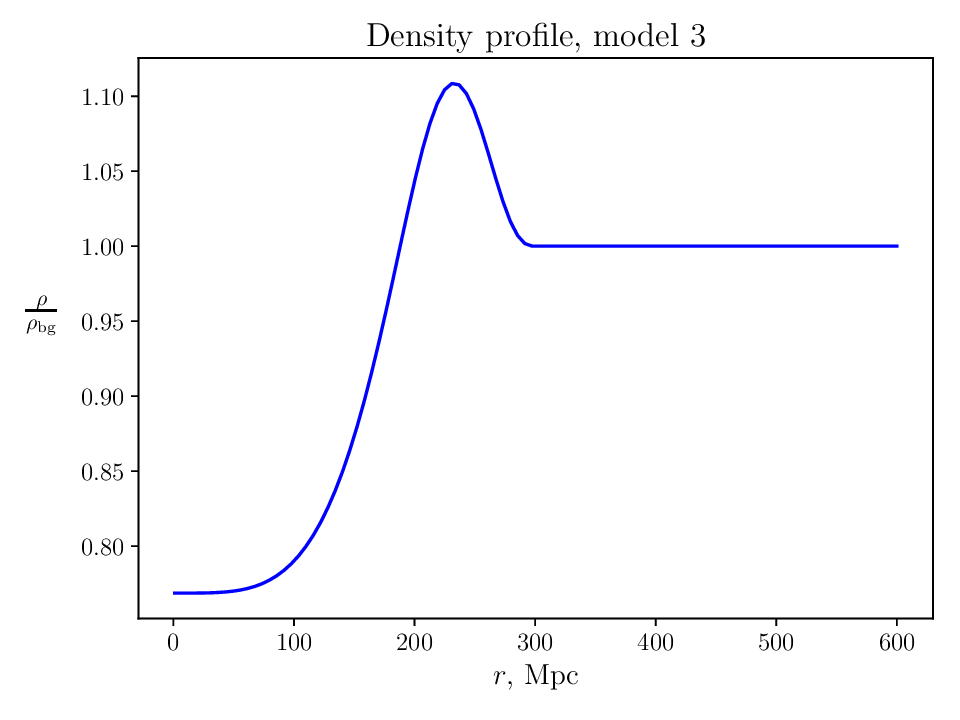}%
	\includegraphics[width=0.45\linewidth]{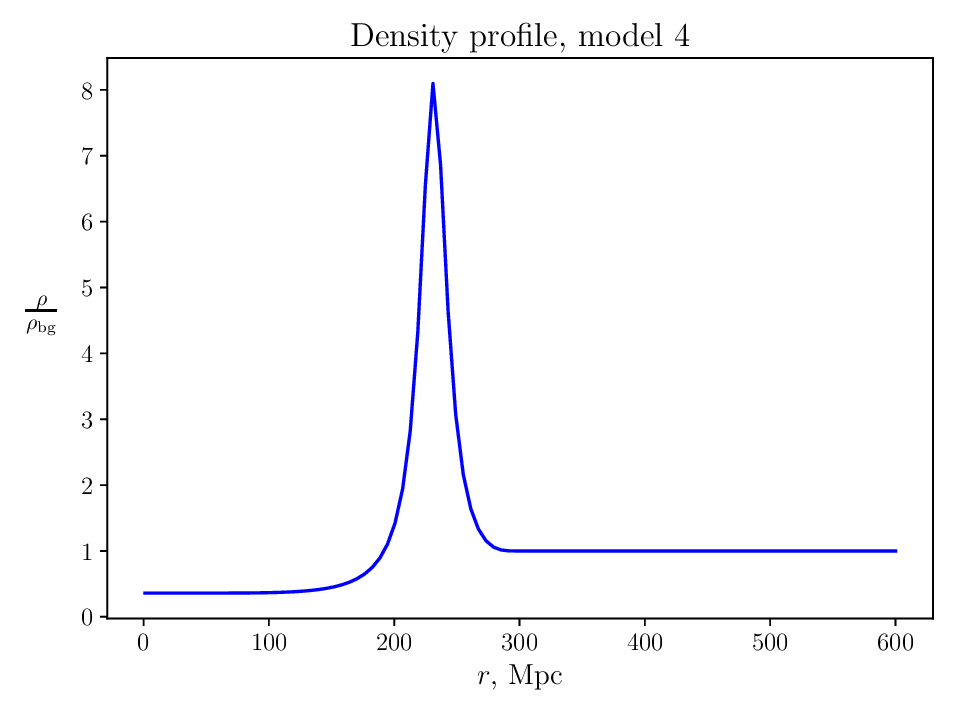}%
	\caption{Present-time density profiles of models 1-4. The densities are shown as fractions of the background density, $\rho_{bg}$.}
	\label{fig:rho_model1}
\end{figure}
\subsection{Specific models}
In this subsection we specify the six different models that we will study, by introducing their $k(r)$. The first four models are motivated by the observational evidence that our local environment is slightly underdense compared to the cosmic mean \cite{local_1, local_2, local_3, local_4, local_5}. These four models thus represent four versions of our local underdensity. The last two models are made to represent The Local Void and its surrounding wall, based on \cite{complexity}.
\newline\indent
\begin{figure}[]    
	\centering
	\includegraphics[width=0.45\linewidth]{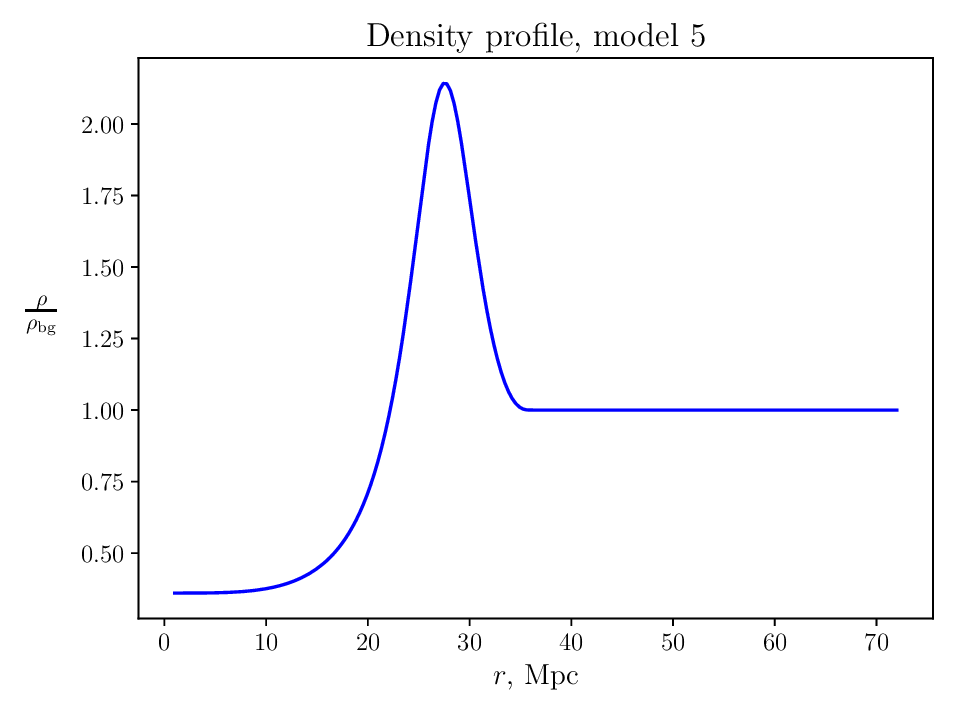}%
	\includegraphics[width=0.45\linewidth]{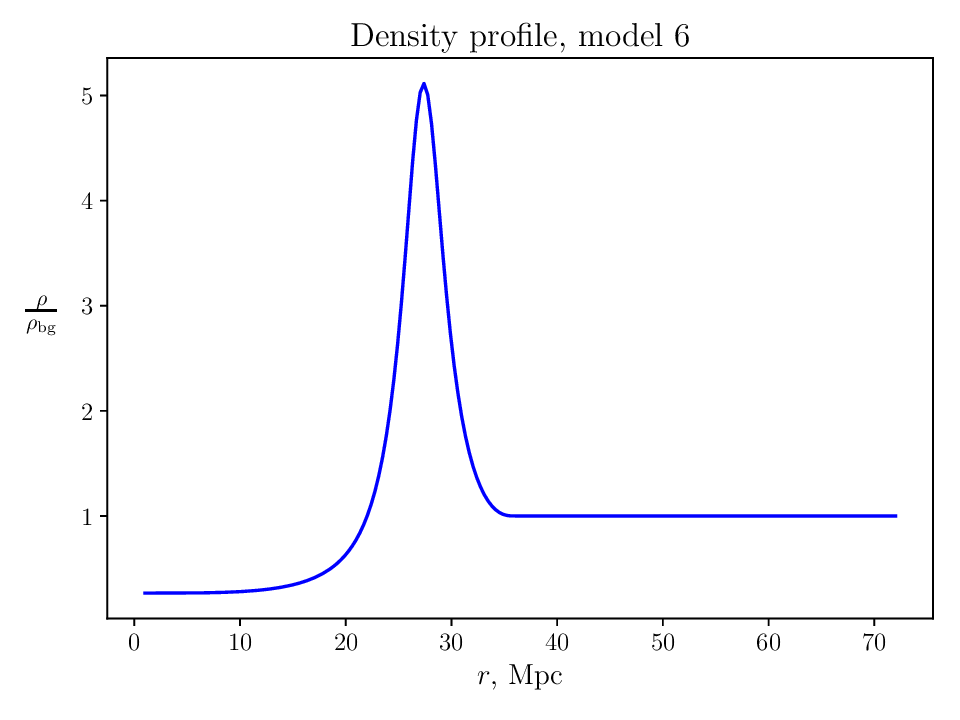}
	\caption{Present-time density profiles of models 5 and 6. The densities are shown as fractions of the background density, $\rho_{bg}$.}
	\label{fig:rho_model5}
\end{figure}
The density profiles presented below are clearly much too simple to faithfully mimic our real local cosmic environment. Nonetheless, being motivated by our real cosmic neighborhood on different scales, we expect that the models can shed light on how we may expect inhomogeneities to affect the convergence and precision of cosmographic expansions relevant for real observations, highlighting possible caveats of cosmographic methods.
\newline\newline
{\bf\underline{Models 1-4: A large local void}}\newline
Observations \cite{local_2} indicate that we are in a slightly underdense region with radius up to around 300Mpc/{\em h} and a local underdensity about 2/3 of the cosmic mean. However, the exact size, shape and depth of this (possible) void is unknown and other observational studies indicate other values for its radius and maximum underdensity \cite{local_1, local_3, local_4}, with the maximum underdensity in the range $4-45\%$ and radius around $\sim 100$Mpc/{\em h}, and with us not necessarily placed particularly near the center (and in \cite{local_5} it was found that signs of a local underdensity depends on data binning). We will therefore consider four void density profiles, all with radius 300Mpc but different profiles and maximum depths. Note that the specific choice of radius does not qualitatively affect our results which is why we only use a single (somewhat arbitrary) radius. The final specification of the four void models are
\begin{align}
\begin{split}
k_1(r) &= -3.9\cdot 10^{-9}\cdot r^2\exp\left(-\left(\frac{r}{r_b/1.2}\right)^2\right)\\
k_2(r) &= -3\cdot 10^{-8}\cdot r^2\exp\left(-\left(\frac{r}{r_b/1.2}\right)^2\right)\\
k_3(r) & = \left\{ \begin{array}{rl}
-10^{-8}r^2\left(\left(\frac{r}{r_b} \right)^4 -1 \right)^4  &\text{if} \,\, r<r_b \\
0 &\mbox{ otherwise.}
\end{array} \right.\\
k_4(r) & = \left\{ \begin{array}{rl}
-5\cdot 10^{-8}r^2\left(\left(\frac{r}{r_b} \right)^6 -1 \right)^6  &\text{if} \,\, r<r_b \\
0 &\mbox{ otherwise.}
\end{array} \right.,
\end{split}
\end{align}
where $r_b = 300$Mpc in all four cases.\newline\indent
The present-time density profiles corresponding to these four versions of $k(r)$ are shown in figure \ref{fig:rho_model1}. A seen, models 1 and 2 represent voids (of various depth) with smooth, asymptotic transitions towards the background FLRW model. The two other models have central voids compensated exactly by surrounding overdensities and reduce exactly to their background FLRW models at $r = r_b$.
\newline\newline
{\bf\underline{Models 4-6: The Local Void and its wall(s)}}
\newline
Through a more thorough morphological description of our near cosmic neighborhood we can describe the Milky Way as placed near the edge of an incomplete wall surrounding ``The Local Void'' (see figure 1 of \cite{complexity} and note that The Local Void is a specific structure, different from the larger possible void/underdensity discussed in relation to models 1-4). Although The Local Void is clearly not spherically symmetric and thus cannot faithfully be traced by an LTB model, we will make a rough approximation of it as an LTB model with present-day radius 36Mpc\footnote{We find this radius by noting that \cite{complexity} estimated the volume of the Local Void to be roughly $2\cdot 10^5\rm Mpc^3$.}. Since The Local Void is not actually spherically symmetric and e.g. has several local minima and maxima, we will vary the depth of the void and height of the overdensity and use two versions of $k(r)$ to achieve this. Specifically, we consider models very similar to models 3 and 4, with 
\begin{align}
k_5(r) & = \left\{ \begin{array}{rl}
-5\cdot 10^{-8}r^2\left(\left(\frac{r}{r_b} \right)^4 -1 \right)^4  &\text{if} \,\, r<r_b \\
0 &\mbox{ otherwise.}
\end{array} \right.\\
k_6(r) & = \left\{ \begin{array}{rl}
-7.2\cdot 10^{-8}r^2\left(\left(\frac{r}{r_b} \right)^4 -1 \right)^4 &\text{if} \,\, r<r_b \\
0 &\mbox{ otherwise.}
\end{array} \right.,
\end{align}
where $r_b = 36$Mpc. The corresponding present-day density profiles are shown in figure \ref{fig:rho_model5}. Note that model 4 only differs from model 5 by $r_b$ which, however, is enough to make the values of their peak densities quite different from each other. Although the difference between models 5 and 6 versus models 3 and 4 are overall modest, we will place observers somewhat differently in these two sets of models and in addition, their different sizes mean that light rays will be computed for different redshift intervals for the two groups of models.

\section{Redshift-distance relations}\label{sec:light}
For studying the accuracy and convergence of cosmographic expansions of the redshift-distance relation, we need to compute the exact redshift-distance relation along light rays in the LTB models for comparison. We summarize the method for this below before moving on to describing the cosmographic expansions we consider.
\newline\newline
We will assume that the geometric optics approximation holds and that light thus moves along null-geodesics which means that the light paths can be traced by solving the geodesic equations
\begin{align}
\frac{d}{d\lambda}\left( g_{\alpha\beta}k^\beta \right) = \frac{1}{2}g_{\beta\gamma,\alpha}k^{\beta}k^{\gamma},
\end{align}
where $\lambda$ is an affine parameter along the null geodesic and $k^\alpha$ is the light ray tangent vector which is initialized with $k^t = -1/c$ and such that $k^\alpha k_\alpha = 0$. We can then compute the redshift along the light ray as $1+z = \left(k^\mu u_\nu\right)_e/\left(k^\mu u_\nu\right)_o$, where subscripts $e$ and $o$ indicate evaluation at the position of emission and observation, respectively.
\newline\indent
Simultaneously with solving the geodesic equations, we solve propagation equations for parallel transporting the screen space basis vectors, $E_1^\mu, E_2^\mu$, along the light ray. The screen space basis vectors are orthogonal unit vectors spanning the space orthogonal to both the observer velocity and $k^\alpha$ (screen space). Lastly, we must (simultaneously) solve the transport equation \cite{light}
\begin{align}
\frac{d^2 }{d\lambda^2}D^a_ b = T^a_c D^c_b, 
\end{align}
where $D^a_b$ is the deformation tensor from which we can obtain the angular diameter distance as $D_A = \sqrt{|\det(D)|}$. To obtain the luminosity distance we simply use the reciprocity relation from which we have $D_L = (1+z)^2D_A$.
\newline\indent
Introducing $\epsilon^\mu := E_1^\mu + iE_2^\mu$, we can write the components of the tidal matrix, $T_{ab}$, as
\begin{equation}\label{eq:tidal}
T_{ab} = 
\begin{pmatrix} \mathbf{R}- Re(\mathbf{F}) & Im(\mathbf{F}) \\ Im(\mathbf{F}) & \mathbf{R}+ Re(\mathbf{F})  \end{pmatrix} .
\end{equation}
We have here introduced $\mathbf{R}:=-\frac{1}{2}R_{\mu\nu}k^{\mu}k^{\nu}$ and $\mathbf{F}:=-\frac{1}{2}R_{\alpha\beta\mu\nu}(\epsilon^*)^\alpha k^\beta (\epsilon^*)^\mu k^\nu$, with $R_{\alpha\beta}$ denoting the Ricci tensor and $R_{\alpha\beta\mu\nu}$ the Riemann tensor of the LTB spacetime.
\newline\newline
By solving the above set of 24 ordinary differential equation simultaneously, we can obtain the exact redshift-distance relation along light rays in LTB models.

\subsection{Cosmographic expansions}
Cosmographic expansions are Taylor expansions applied to cosmologically relevant quantities such as the Hubble parameter, dark energy equation of state parameter and redshift-distance relations. Early considerations were focused on FLRW spacetimes, where the three lowest order expansion coefficients of the scale factor are
\begin{align}
\begin{split}
H(t) &:= \frac{\dot a}{a}\\
q(t) &:= -\frac{\ddot a}{aH^2}\\
j(t) & := \frac{\dddot a}{aH^3}.
\end{split}
\end{align}
These are known as the Hubble, deceleration and jerk parameter, respectively. (See e.g. \cite{jerk_1, jerk_2} for details). However, it is possible to generalize the cosmographic expansions to more complicated spacetimes. We will here thus also consider the series expansion for the luminosity distance presented in \cite{general_1}.
\newline\indent
According to \cite{general_1}, the general cosmographic expansion of the luminosity distance is given by (assuming that the redshift is monotonous along the light ray)
\begin{align}\label{eq:DL}
D_L \approx D_L^{(1)}z + D_L^{(2)}z^2 + D_L^{(3)}z^3,
\end{align}
with
\begin{align}
\begin{split}
D_L^{(1)} &= \frac{c}{\mathcal{H_O}}\\
D_L^{(2)} & = \cfrac{1-\mathcal{Q_O}}{2\mathcal{H_O}}\\
D_L^{(3)} & = c\frac{-1 + 3\mathcal{Q_O}^2+ \mathcal{Q_O - \mathcal{J_O}+\mathcal{R_O}}}{6\mathcal{H_O}}.
\end{split}
\end{align}
These expressions include generalized versions of spatial curvature and the Hubble, deceleration and jerk parameters given by (ignoring acceleration of the dust of our spacetime since this vanishes in the LTB models)
\begin{align}\label{eq:coeff}
\begin{split}
\mathcal{H_O} &= \left(\frac{1}{3}\Theta +e^\mu e^\nu \sigma_{\mu\nu}\right)|_{\mathcal{O}}\\
\mathcal{Q_O} & = -1 - \frac{c^2}{E_\mathcal{O}\mathcal{H_O}^2}\frac{d\mathcal{H}}{d\lambda}|_{\mathcal{O}}\\
\mathcal{R_O} & = 1+\mathcal{Q_O} - \frac{c^4}{2E_{\mathcal{O}}}\frac{k^\mu k^\nu R_{\mu\nu}|_{\mathcal{O}}}{\mathcal{H_O}^2}\\
\mathcal{J_O} & = \frac{c^4}{E_{\mathcal{O}}^2\mathcal{H_O}^3}\frac{d^2\mathcal{H}}{d\lambda^2}|_{\mathcal{O}}-4\mathcal{Q_O}- 3,
\end{split}
\end{align}
where $e^{\mu} = u^{\mu} - ck^\mu/(-u_{\alpha}k^{\alpha})$ is the spatial direction vector of the light ray as seen by an observer comoving with the dust, $\sigma_{\mu\nu}$ is the shear tensor of the dust and $\Theta$ its expansion rate, and $E_{\mathcal{O}}$ is the energy of the photon in the dust frame, $E = -u^\alpha k_{\alpha} $, evaluated at the observer.
\newline\indent
We will also consider the naive isotropized extrapolations of FLRW expressions discussed in \cite{general_1},
\begin{align}
\begin{split}
\mathcal{H}_{\rm naive} &= \Theta/3\\
\mathcal{Q}_{\rm naive} & = -1 -3\frac{d\Theta}{dt}\frac{1}{\Theta^2}\\
\mathcal{R}_{\rm naive} & = -3/2^{(3)}\mathcal{R}/\Theta^2\\
\mathcal{J}_{\rm naive} & = 1+ 9\left( \frac{d^2\Theta}{dt^2} + \Theta \frac{d\Theta}{dt} \right)/\Theta^3,
\end{split}
\end{align}
where $^{(3)}\mathcal{R}$ is the spatial curvature of the LTB model evaluated at the observer, and $\Theta$ is the local expansion rate of the LTB model. We have here omitted subscripts $\mathcal{O}$ to avoid cluttered notation, but for the cosmographic expansions, all quantities above are evaluated at the observer. Using these coefficients in Eq. \ref{eq:coeff} renders the expansion in Eq. \ref{eq:DL} equivalent to the FLRW expression, but where coefficients are evaluated locally at the observer position of the LTB model. 
\newline\newline
In the next section we will present results based on propagating light rays in the six LTB models specified earlier and computing the luminosity distance using the exact relation and the three cosmographic expansions discussed above. However, before moving on to this, we include a small subsection below detailing some considerations regarding the necessary computations.

\subsubsection{Considerations for numerical light propagation and cosmographic expansions in LTB models}
We solve the necessary ODEs using an embedded Runge-Kutta Cash-Karp method implemented with the GNU Scientific Library (GSL)\footnote{https://www.gnu.org/software/gsl/}. The ODEs require high order derivatives of the metric functions which we obtain analytically e.g. by differentiating equation \ref{eq:dynamical}. We evaluate the null-condition, $k^\alpha k_\alpha = 0$, along the light rays to confirm the correctness and precision of our computations.
\newline\indent
The coefficients for the cosmographic expansions are straightforward to obtain with the exception of the derivatives along the light ray, $d\mathcal{H}/d\lambda|_\mathcal{O}$ and $d^2\mathcal{H}/d^2\lambda|_{\mathcal{O}}$. We compute these numerically after having propagated the light rays a few steps in the affine parameter. For this we use fourth order forward finite differences (comparing with second order finite differences and FLRW limits to confirm correctness). For the first order derivative we also compare with the analytical expression. To obtain this, we first note that for the LTB models, we have
\begin{align}
\Theta =\frac{A_{,tr}}{A_{,r}} + 2\frac{A_{,t}}{A}
\end{align}
and
\begin{align}
\sigma^{\mu}_{\nu} = \rm diag\left(0 , -\frac{2}{3}, \frac{1}{3}, \frac{1}{3}\right)\cdot \left( \frac{A_{,t}}{A} - \frac{A_{,tr}}{A_{,r}}\right).
\end{align}
We can then obtain $d\mathcal{H}/d\lambda|_\mathcal{O}$ by first writing out
\begin{align}
\begin{split}
\frac{d\mathcal{H}}{d\lambda} = \frac{1}{3}\frac{d\Theta}{d\lambda} + e^{\mu}e^{\nu}\frac{d\sigma_{\mu\nu}}{d\lambda} + 2\sigma_{ij}e^i \frac{de^j}{d\lambda}\delta_{ij},
\end{split}
\end{align}
where we utilized that $\sigma_{\mu\nu}$ for the LTB model is diagonal with vanishing $tt$-component. Since $\frac{d}{d\lambda} = k^\alpha \partial_{\alpha}$, the necessary derivatives are
\begin{align}
\begin{split}
\frac{d\Theta}{d\lambda} &= k^t \Theta_{,t} + k^r \Theta_{,r}\\
& = k^t\left( \frac{A_{,ttr}}{A_{,r}} - \frac{A_{,tr}^2}{A_{,r}^2}+ 2\frac{A_{,tt}}{A} - 2\frac{A_{,t}^2}{A^2}\right)\\
&+ k^r\left( \frac{A_{,trr}}{A_{,r}} - \frac{A_{,tr}A_{,rr}}{A_{,r}^2} + 2\frac{A_{,tr}}{A} - 2\frac{A_{,t}A_{,r}}{A^2}\right),
\end{split}
\end{align}
\begin{align}
\begin{split}
\frac{d\sigma_{i}^i}{d\lambda} &\propto \frac{d}{d\lambda}\left( \frac{A_{,t}}{A} - \frac{A_{,tr}}{A_{,r}}\right) \\ &=
k^t\left( \frac{A_{,tt}}{A} - \frac{A_{,t}^2}{A^2} - \frac{A_{,ttr}}{A_{,r}} - \frac{A_{,tr}^2}{A_{,r}^2}\right)\\
&+ k^r\left( \frac{A_{,tr}}{A} - \frac{A_{,t}A_{,r}}{A^2} - \frac{A_{,trr}}{A_{,r}} + \frac{A_{,tr}A_{,rr}}{A_{,r}^2} \right),
\end{split}
\end{align}
where a sum is {\em not} implied over $i$ and where we use $\propto$ rather than $=$ since we need prefactors of $1/3, -2/3$, depending on $i = r,\theta, \phi$. We lastly need
\begin{align}
\begin{split}
\frac{de^i}{d\lambda}&= \frac{1}{c}\frac{d}{d\lambda}\frac{k^i}{k^t} \\
&= \frac{1}{ck^t} \frac{dk^i}{d\lambda} - \frac{k^i}{c (k^t)^2}\frac{dk^t}{d\lambda}.
\end{split}
\end{align}

\begin{table}[!htb]
	\centering
	\begin{tabular}{c c c c c c c c}
		\hline\hline
		Model &  $\mathcal{H_O}$ (km/s/Mpc) &  $\mathcal{Q_O}$ & $\mathcal{Q}_{\rm naive}$ & $\mathcal{R_O}$ &$\mathcal{R}_{\rm naive}$ & $\mathcal{J_O}$& $\mathcal{J}_{\rm naive}$\\
		\hline
		\Tstrut
		Model 1, obs1 & 71.3  & -0.511 & -0.546 & 0.0986 & 0.0641 &  -24.9  & 0.936   \\
		Model 1, obs2 & 70.8 & -0.0477 & -0.547 & 0.547 & -0.0501 & -11.9 & 0.950   \\
		Model 1, obs3 & 69.9 & -0.0690 & -0.549 & 0.495 & 0.0208 & 14.3  & 0.979\\
		Model 2, obs1 & 76.7 & -0.430 & -0.521 & 0.381 & 0.290 & -51.5 & 0.710\\
		Model 2, obs2 & 74.7 & 1.125 & -0.527 & 2.02  & 0.250 & -66.9 & 0.751\\
		Model 2, obs3 & 69.4 & 3.08 & -0.540& 3.72 & 0.128 & 16.2 & 0.873\\
		Model 3, obs1 & 72.9  & -0.539  & -0.539  & 0.142   & 0.142  & 0.234   & 0.858 \\
		Model 3, obs2 & 72.4  & 0.622 & -0.540 & 1.29 & 1.29 & -135 & 0.871 \\
		Model 3, obs3 & 66.2 & 4.93 & -0.551 & 5.41 & -0.0302 & 566  & 1.03\\
		Model 4, obs1 & 79.0  & -0.508  & -0.508  & 0.365   & 0.365  & 0.627   & 0.365 \\
		Model 4, obs2 & 78.7 & 0.00650 &  -0.508&  0.876&  0.362 & -72.3 & 0.638 \\
		Model 4, obs3 &  41.0 & 858 & -0.516 & 857 & -0.404& $-493\cdot 10^4$  & 1.45\\
		FLRW limit & 70.0 & -0.55  & -& 0 & - & 1 & - \\
		\hline
	\end{tabular}
	\caption{Cosmographic expansion parameters for models 1-4 for the three observers together with the FLRW limit of the considered LTB models.}
	\label{table:Models1-3}
\end{table}

\section{Results}\label{sec:results}
\begin{figure}[]
	\centering
	\includegraphics[width=0.45\linewidth]{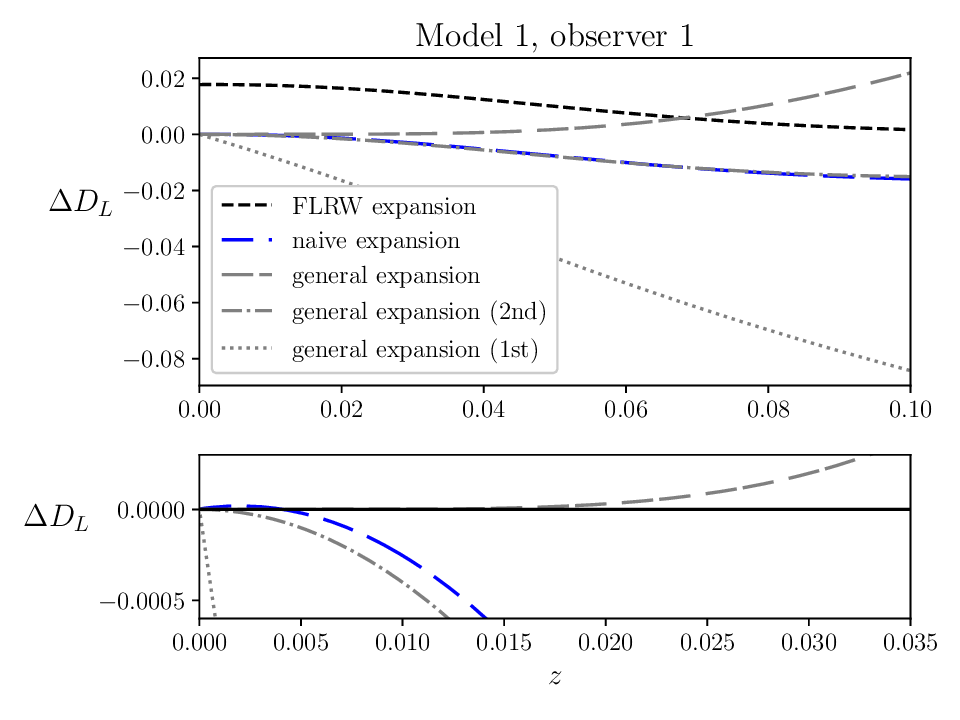}
	\includegraphics[width=0.45\linewidth]{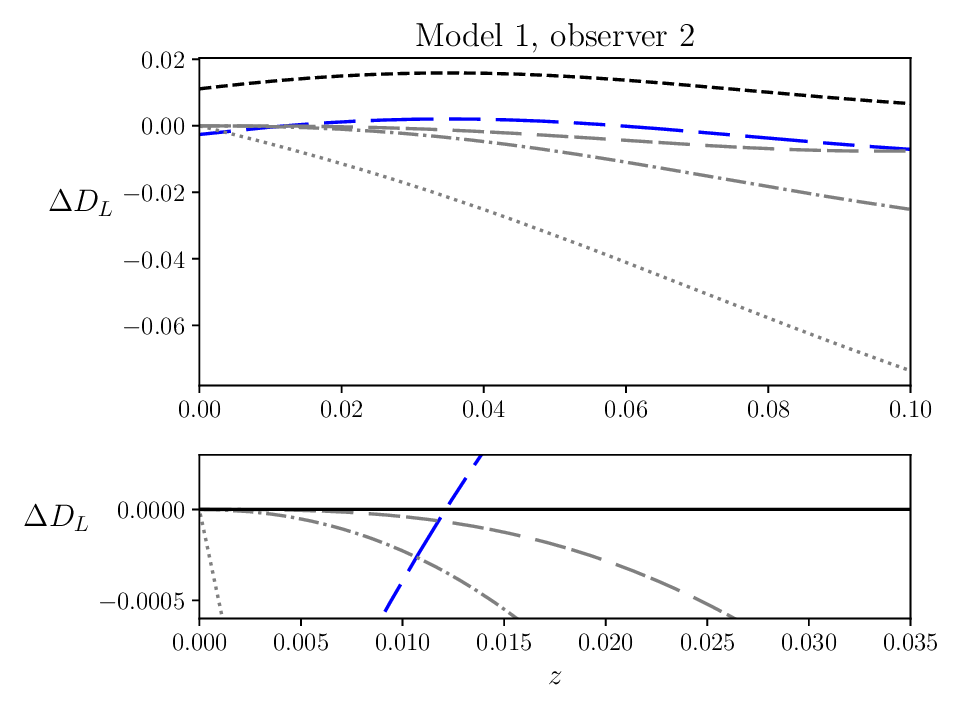}
	\includegraphics[width=0.45\linewidth]{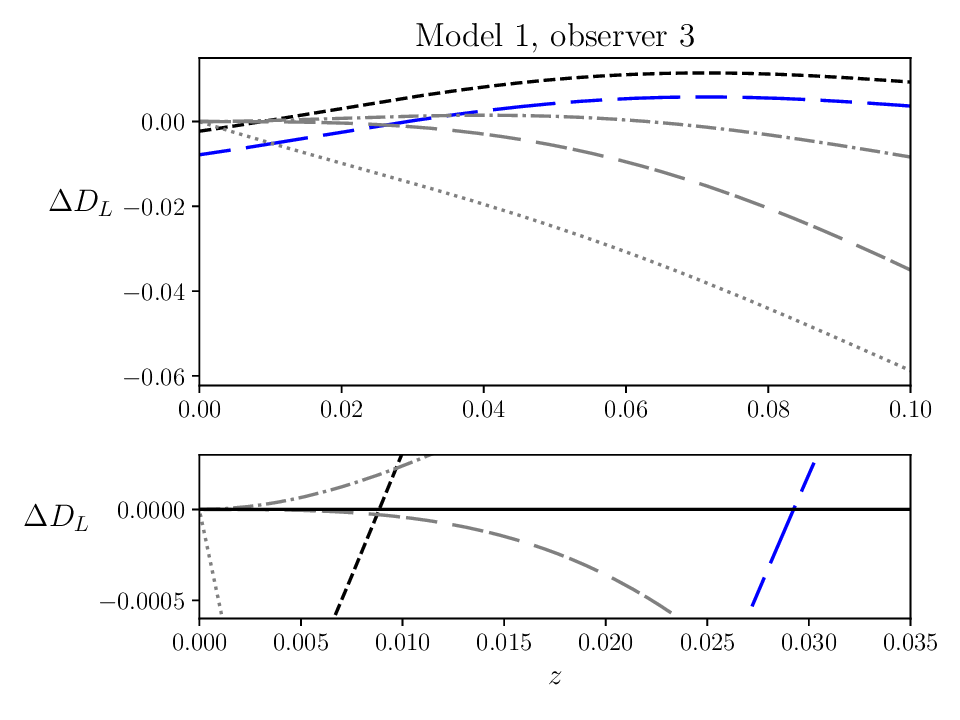}
	\caption{Cosmographic expansions shown as relative deviations from the exact luminosity distance, $\Delta D_L :=({\rm cosmographic\,\,\, expansion}-D_L)/D_L$, along fiducial lines of sight for three present-time observers in model 1. The naive and FLRW cosmographic expansions are only shown at third order. The general expansion is shown at first, second and third order. An inset with a black zero-line is included to show the low-z behavior of the expansions.}
	\label{Fig:ddl_model1}
\end{figure}

\begin{figure}[]
	\centering
	\includegraphics[width=0.45\linewidth]{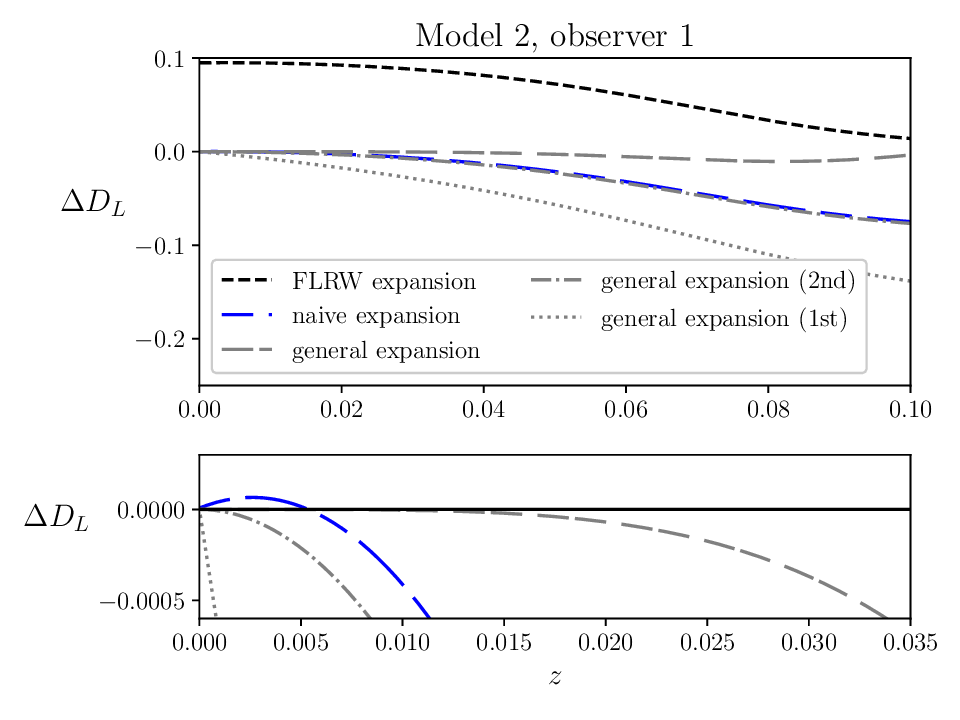}
	\includegraphics[width=0.45\linewidth]{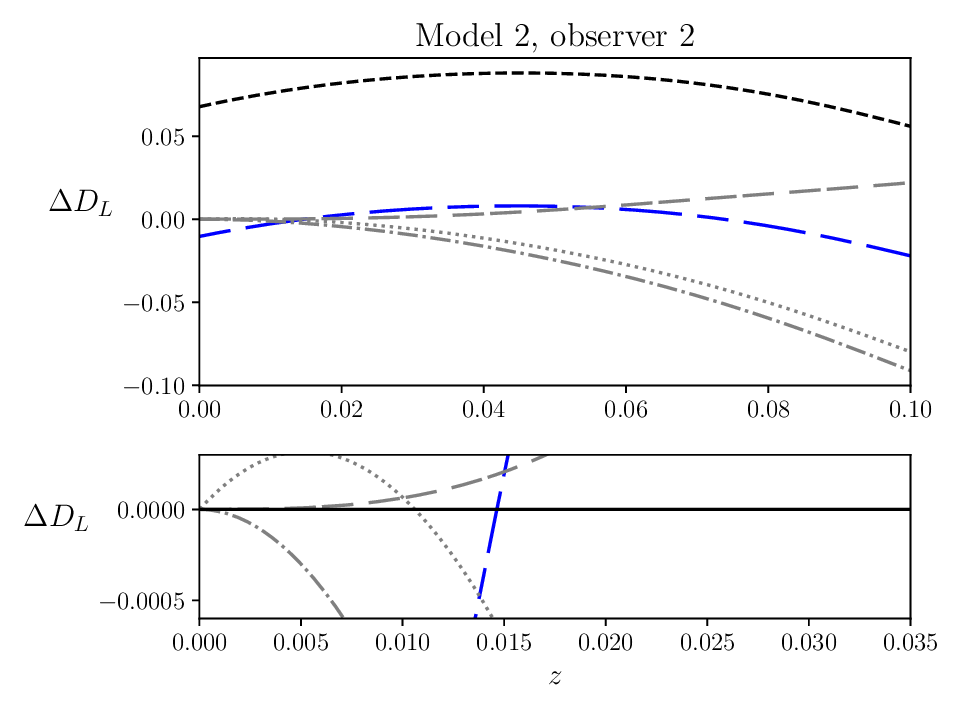}
	\includegraphics[width=0.45\linewidth]{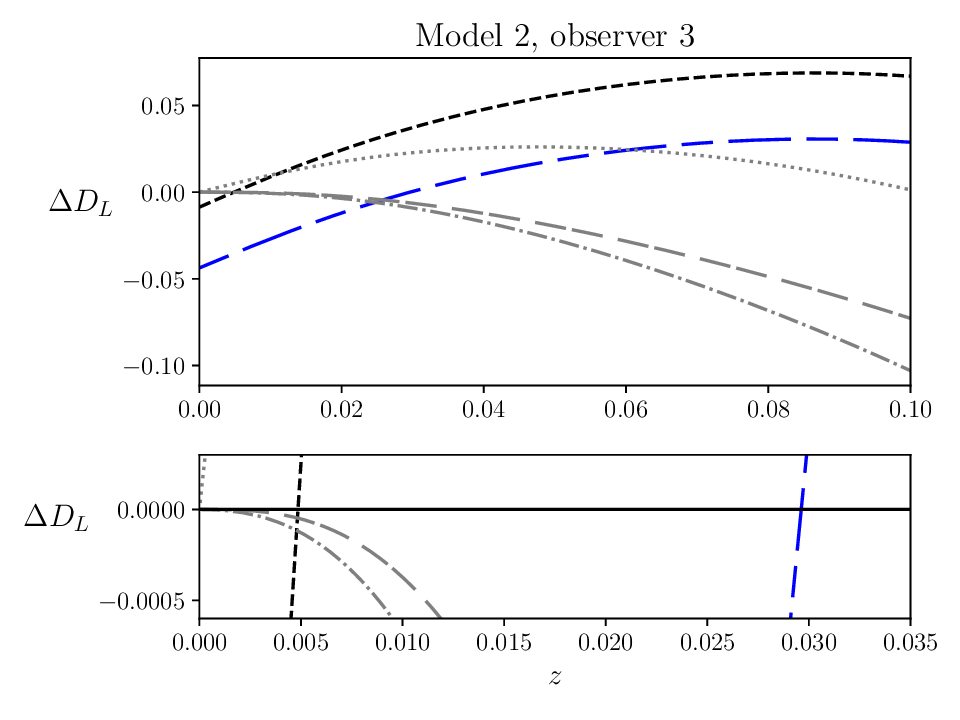}
	\caption{Cosmographic expansions shown as relative deviation from the exact luminosity distance, $\Delta D_L :=({\rm cosmographic\,\,\, expansion}-D_L)/D_L$, along fiducial lines of sight for three present-time observers in model 2. The naive and FLRW cosmographic expansions are only shown at third order. The general expansion is shown at first, second and third order. An inset with a black zero-line is included to show the low-z behavior of the expansions.}
	\label{Fig:ddl_model2}
\end{figure}

\begin{figure}[]
	\centering
	\includegraphics[width=0.45\linewidth]{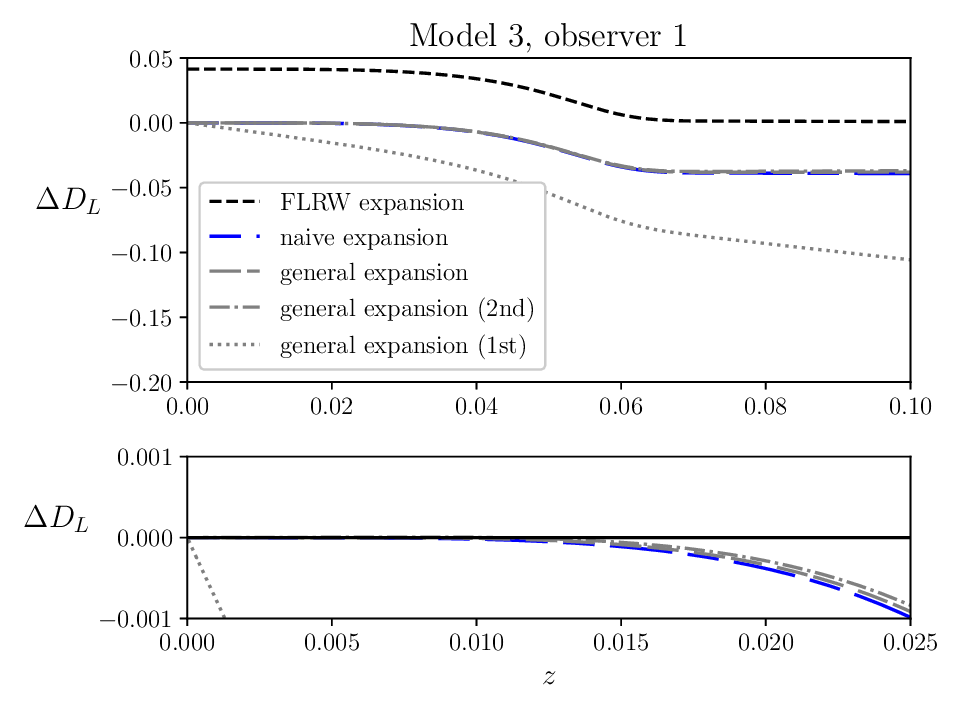}
	\includegraphics[width=0.45\linewidth]{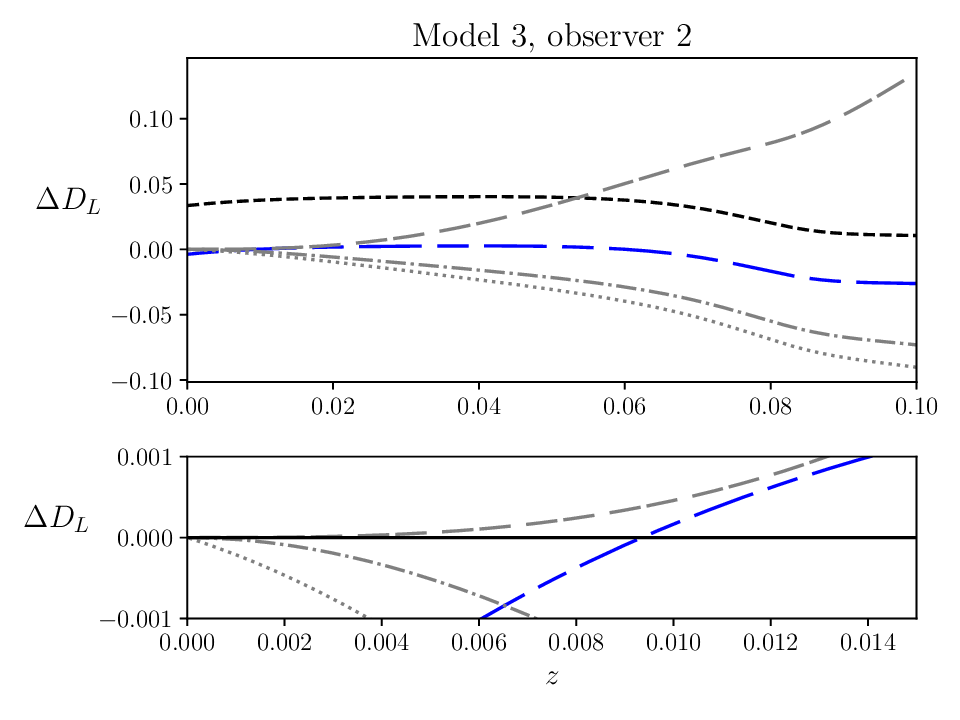}
	\includegraphics[width=0.45\linewidth]{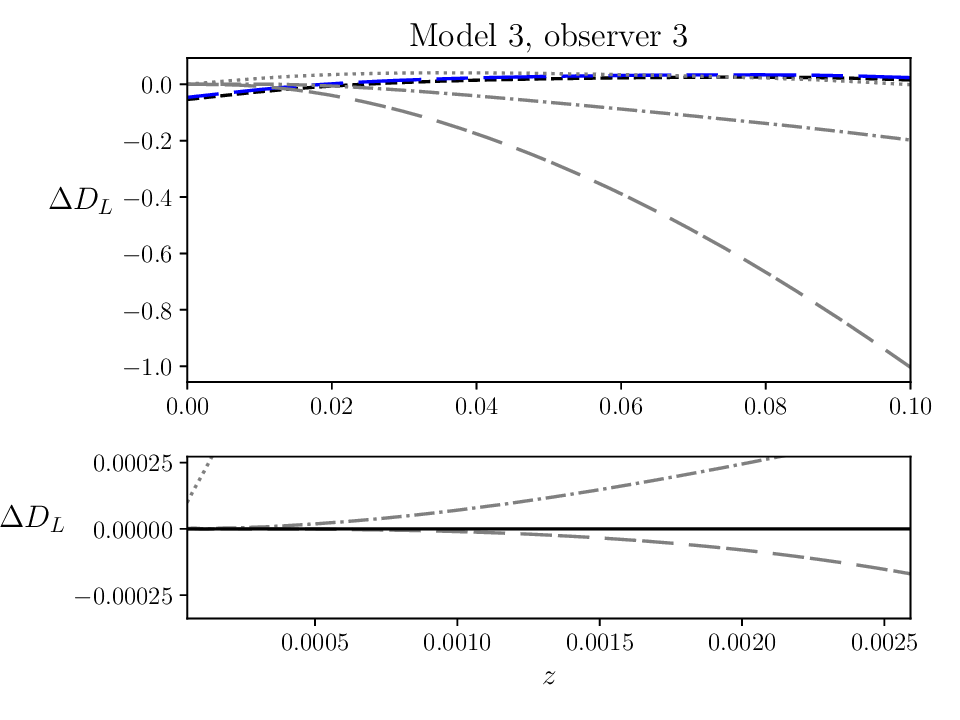}
	\caption{Cosmographic expansions shown as relative deviation from the exact luminosity distance, $\Delta D_L :=({\rm cosmographic\,\,\, expansion}-D_L)/D_L$, along fiducial lines of sight for three present-time observers in model 3. The naive and FLRW cosmographic expansions are only shown at third order. The general expansion is shown at first, second and third order. An inset with a black zero-line is included to show the low-z behavior of the expansions.}
	\label{Fig:ddl_model3}
\end{figure}

\begin{figure}[]
	\centering
	\includegraphics[width=0.45\linewidth]{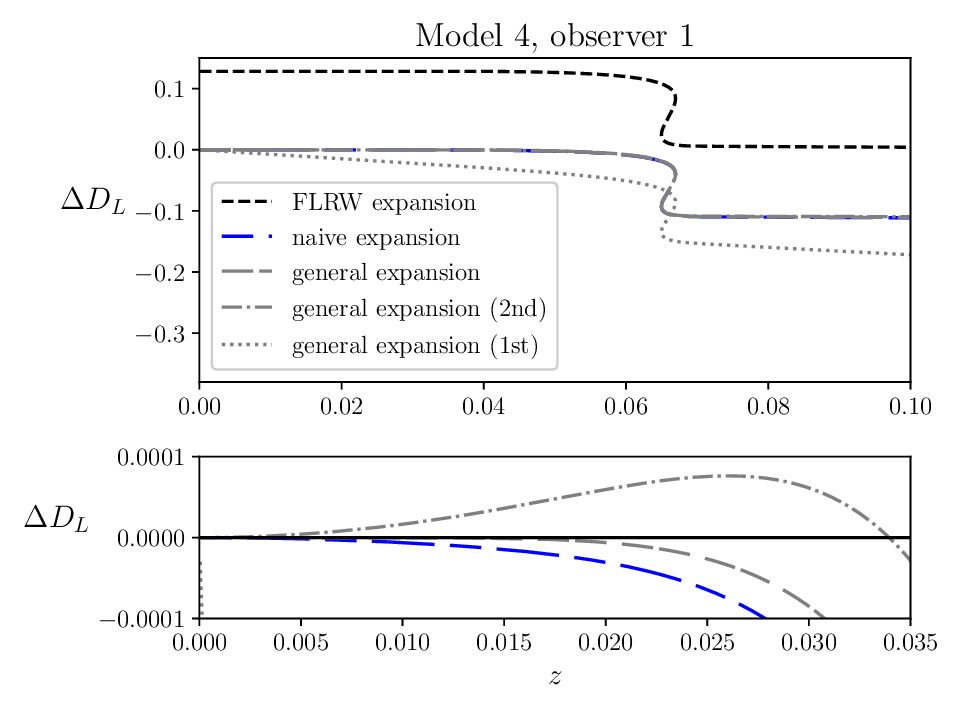}
	\includegraphics[width=0.45\linewidth]{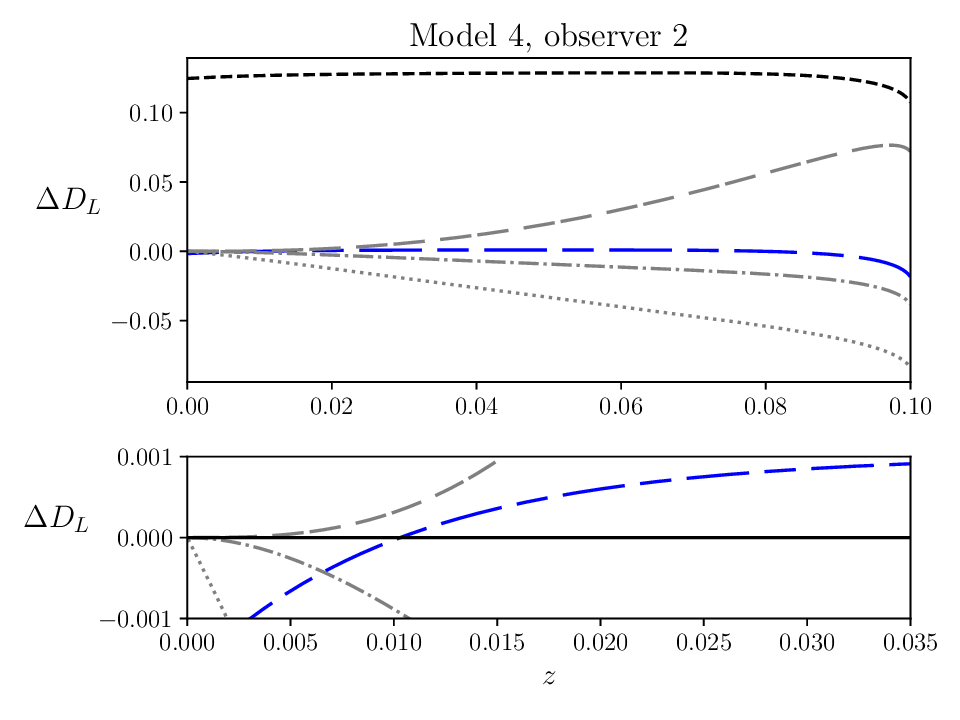}
	\includegraphics[width=0.45\linewidth]{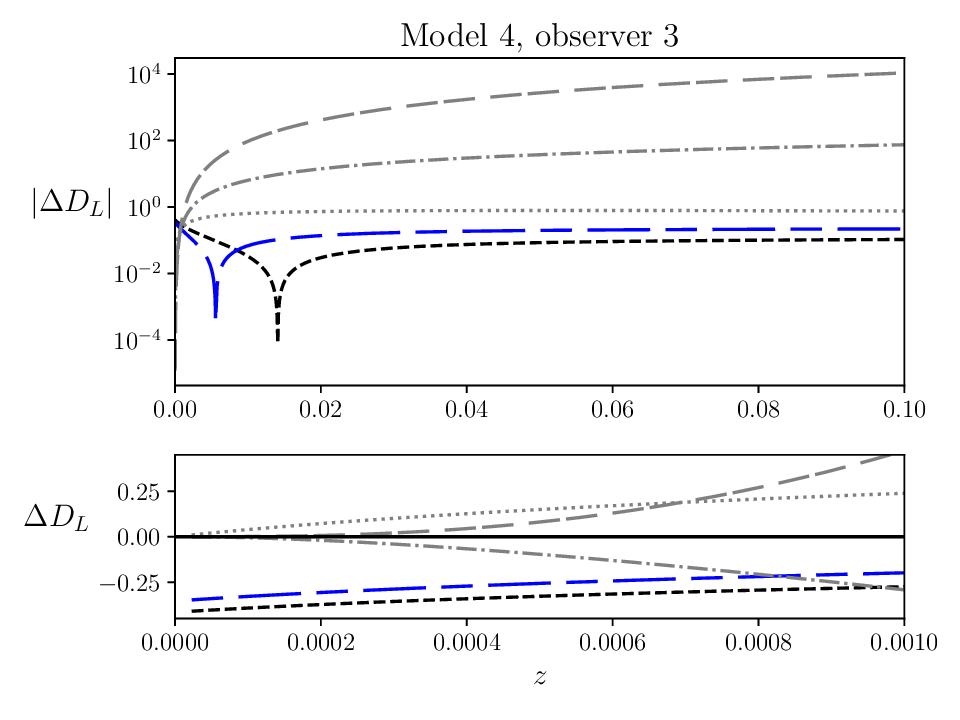}
	\caption{Cosmographic expansions shown as relative deviation from the exact luminosity distance, $\Delta D_L := ({\rm cosmographic\,\,\, expansion}-D_L)/D_L$, along fiducial lines of sight for three present-time observers in model 4. The naive and FLRW cosmographic expansions are only shown at third order. The general expansion is shown at first, second and third order. An inset with a black zero-line to show the low-z behavior of the expansions.}
	\label{Fig:ddl_model4}
\end{figure}

In this section, we present results from computing the redshift-distance relation along fiducial light rays for observers placed at different distances from the symmetry center in the six LTB models specified earlier. Note that if we place the observer outside the structure, in the (asymptotic) FLRW region, the general cosmographic expansion reduces to the ordinary FLRW-based expansion. Observers are therefore always placed inside the inhomogeneous regions. We first consider a small number of individual (fiducial) lines of sight for all models before moving on to considering a larger number of random lines of sight for two of the models.
\newline\newline
{\bf\underline{Models 1-4: Living in a void}}\newline
For each model, we consider 3 different present-time observers placed at $r = 10$Mpc, $r = 100$Mpc and $r = 200$Mpc, and denote the corresponding observers as observer 1, 2 and 3, respectively. We consider 3 random lines of sight for each observer, though always with $k^r<0$. Table \ref{table:Models1-3} shows the corresponding values of $\mathcal{H_O}, \mathcal{Q_O}, \mathcal{R_O}, \mathcal{J_O}$ for each of the observers and lines of sight together with the corresponding values of the FLRW background. Since the values of $\mathcal{Q_O}, \mathcal{J_O}, \mathcal{R_O}$ and their naive counterparts tend to be very different, we also show the naive extrapolations of the FLRW quantities in the table. Among the more notable points to draw from the table is that $\mathcal{Q_O}$ can have both signs. The naive value of $\mathcal{Q_O}$ is, however, negative for all models and observers studied in this section but we note that we in the next section find observes and models where the naive extrapolation of $\mathcal{Q_O}$ is positive as well. Overall, we note that the local environment can significantly alter the coefficients of the cosmographic expansion. Table \ref{table:Models1-3} shows that $\mathcal{Q_O}$ can be orders of magnitude larger or smaller than the naive extrapolation and the $\Lambda$CDM background value, and that $\mathcal{R_O}$ can also deviate from the naive extrapolation by orders of magnitudes (the background value is 0). The general $\mathcal{R_O}$ and its naive FLRW extrapolation do not always have the same sign and this is also true for $\mathcal{J_O}$. We also note that $\mathcal{Q_O}$ and $\mathcal{R_O}$ are almost identical for observer 3 in model 4. This is because the latter actually contains the former as a term in its definition and this term, in this case, is indeed the dominating term of $\mathcal{R_O}$. $\mathcal{J_O}$ also contains a term with $\mathcal{Q_O}$ but the term with $d^2\mathcal{H}/d\lambda^2$ generally dominates $\mathcal{J_O}$ so that $\mathcal{Q_O}$ and $\mathcal{J_O}$ take very different values from each other.
\newline\indent
For each observer and their three random lines of sight, we propagate light rays up to $z = 0.1$. The resulting cosmographic expansions are shown in figures \ref{Fig:ddl_model1}, \ref{Fig:ddl_model2}, \ref{Fig:ddl_model3} and \ref{Fig:ddl_model4} relative to the exact redshift-distance relation. The figures show the general cosmographic expansion at first, second and third order in comparison with the third order FLRW and naive expansions. As demonstrated with insets in the figures we find (as expected) that along all light rays and all models, the third order general cosmographic expansion is the most precise at very low redshift, i.e. close to the observer. On the other hand, the third order cosmographic approximation is rarely the most accurate of the expansions near the end of the studied redshift interval which indicates that $z = 0.1$ is above the radius of convergence of the underlying Taylor series \footnote{We remind the reader that a Taylor series can be used to approximate the underlying function {\em only} within the series' radius of convergence. Once outside the radius of convergence of the Taylor series, the expansion is no longer a good approximation of the function, even if the expansion parameter (here $z$) is much smaller than 1. The radius of convergence can be formally computed as $\lim_{n\rightarrow \infty} |\frac{c_n}{c_{n+1}}|$, where $c_n$ are the expansion coefficients. We do not consider such a formal computation of the radius of convergence here since we only compute 3 coefficients for each expansion and these do not generally have the same sign, making it infeasible to assess the limit. From the divergence of the third order general cosmographic expansion we can nonetheless conclude that the radius of convergence of the expansion has been exceeded along the considered light rays.}. This is especially prominent for observer 3 in model 4 where the third order cosmographic expansion diverges already after a very short distance from the observer, but for almost all models and observers, the general cosmographic expansion clearly begins to diverge before $z = 0.1$ is reached.
\newline\indent

\begin{table}[!htb]
	\centering
	\begin{tabular}{c c }
		\hline\hline
		Model &  radius of convergence ($z$) \\ 
		\hline
		\Tstrut
		Model 1, obs1 & $0.1-1$   \\
		Model 1, obs2 &    $0.1-1$ \\
		Model 1, obs3 &  $0.1-1$\\
		Model 2, obs1 &  $0.1-1$\\
		Model 2, obs2 &  $0.01-1$\\
		Model 2, obs3 &  $0.1-1$\\
		Model 3, obs1 & 1 \\
		Model 3, obs2 &  $0.01-1$ \\
		Model 3, obs3 &  1\\
		Model 4, obs1 &   1\\
		Model 4, obs2 &  $0.1-1$ \\
		Model 4, obs3 & $10^{-3}-10^{-2}$ \\
		FLRW limit & 1 \\
		\hline
	\end{tabular}
	\caption{Estimated order of magnitude of the radius of convergence for models 1-4 for the three observers together with the FLRW background.}
	\label{table:convergence1-3}
\end{table}

\begin{figure}[]
	\centering
	\includegraphics[width=0.45\linewidth]{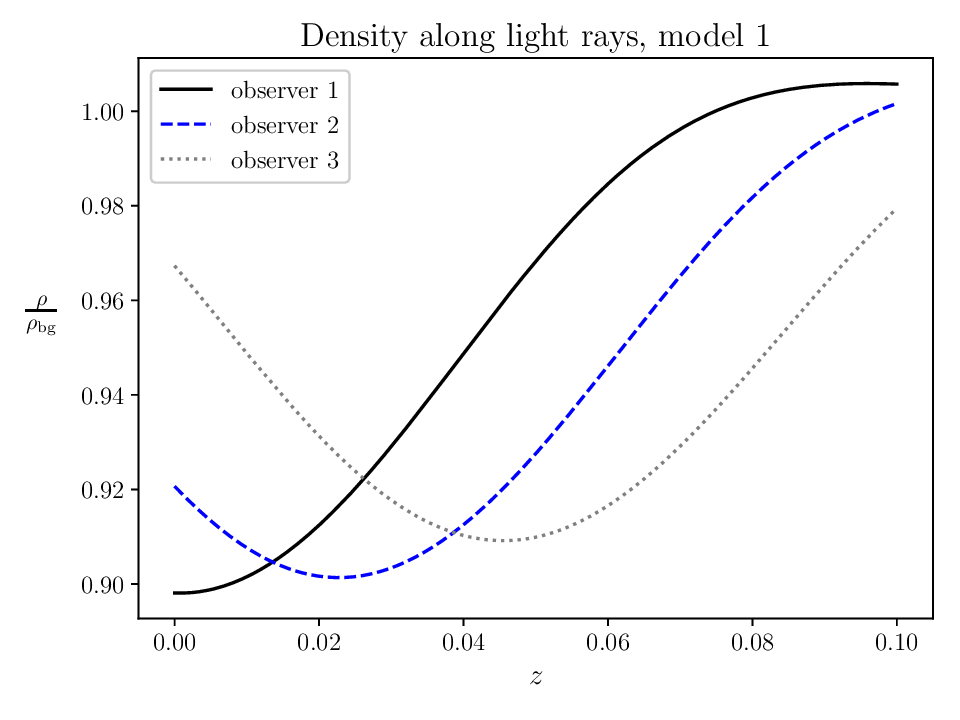}
	\includegraphics[width=0.45\linewidth]{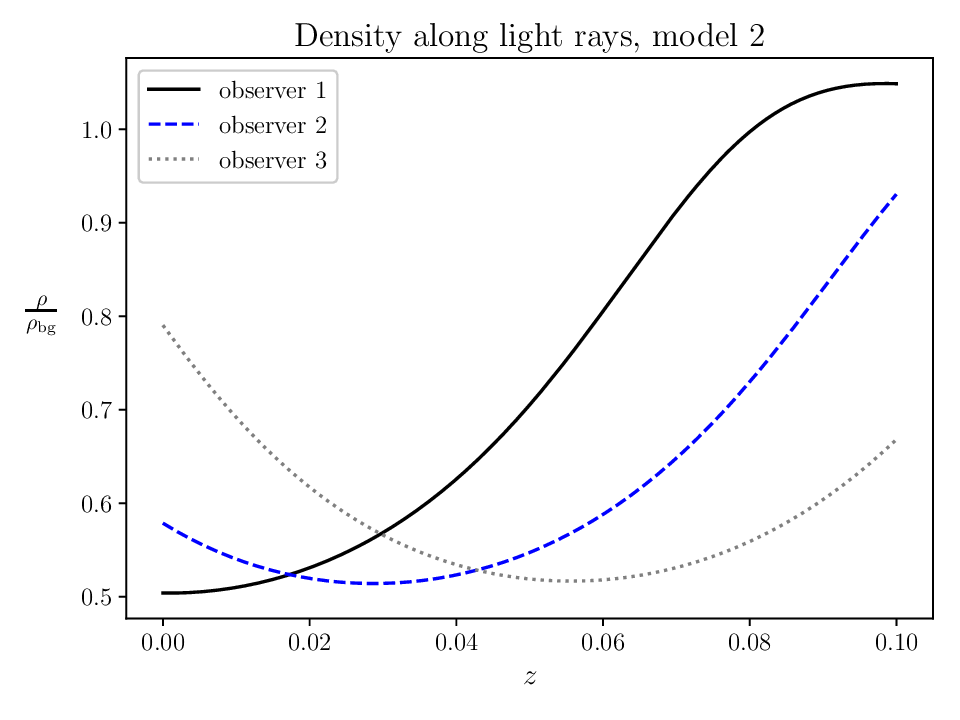}\\
	\includegraphics[width=0.45\linewidth]{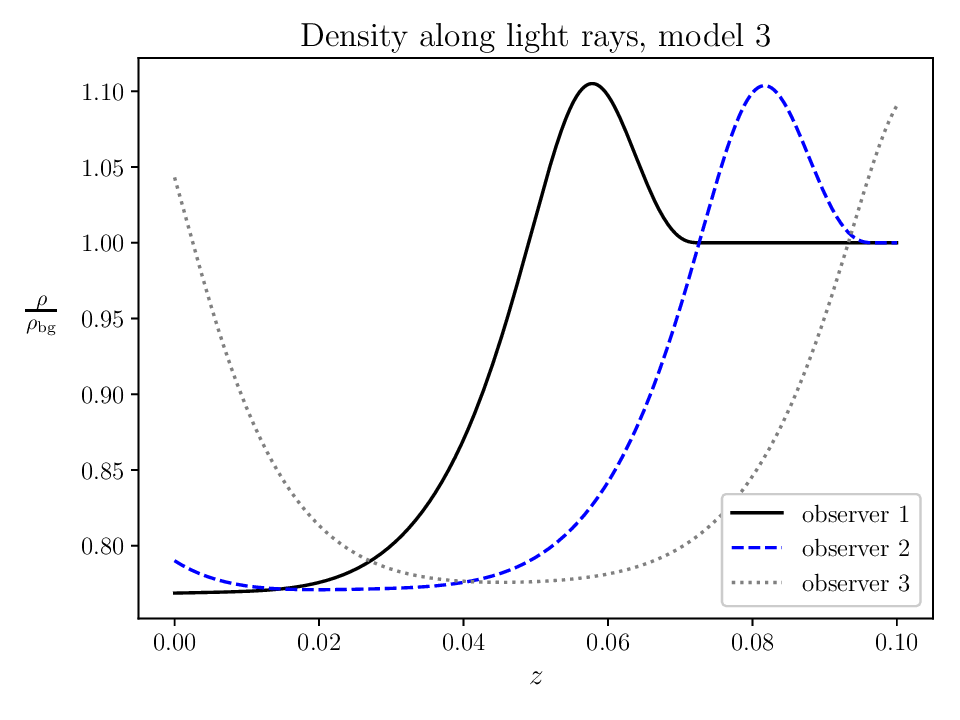}
	\includegraphics[width=0.45\linewidth]{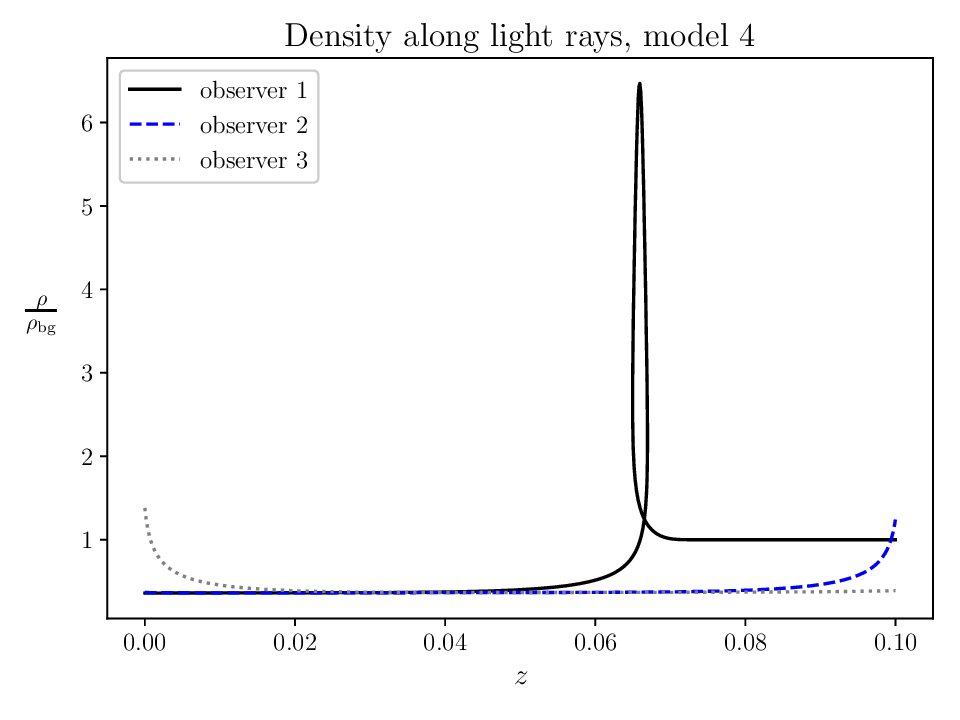}
	\caption{Density along the light rays studied for models 1-4.}
	\label{Fig:ray_rho}
\end{figure}
We further note that when the observer is placed in a locally fairly homogeneous patch (e.g. close to the center of symmetry), the general and naive expansions are very similar because the naive extrapolations of the expansion coefficients are close to the genuine expansion coefficients. For these observers, only the FLRW-based cosmographic expansion is noticeably poor. When the observer is placed further from the center of symmetry, the anisotropy/shear becomes more important, and the naive expansion also becomes clearly inferior to the general expansion near the observer.
\newline\indent
For especially observer 1 but also observer 2, the FLRW cosmographic expansion becomes better at higher redshifts. This can be understood by noting that near $z = 0.1$, the corresponding light rays have all reached spacetimes close to or exactly equal to the FLRW background of the models. This is seen in figure \ref{Fig:ray_rho} which shows the density along each of the considered light rays. When a light ray propagates through an FLRW region, the redshift-distance relation of the LTB models is very close to that of the background FLRW model even after the light ray has traversed one or more inhomogeneities for the models studied here (see e.g. figure 2 in \cite{close} for demonstrations of this, but note that while this claim is true for the considered and similar models, it is not necessarily true for a general LTB model -- a counterexample can be found in e.g. \cite{not_FLRW}).
\newline\indent
By comparing the density along light rays (figure \ref{fig:rho_model1}) with figures \ref{Fig:ddl_model1}-\ref{Fig:ddl_model4} we see that the radius of convergence of the cosmographic expansions seems to depend on the steepness of the density gradient near the observer, which is where $\frac{d\mathcal{H}}{d\lambda}$ and $\frac{d^2\mathcal{H}}{d\lambda^2}$ also tend to be largest and deviate the most from both the naive and FLRW limits. The main difference between $\mathcal{H}$ and its naive FLRW extrapolation is the shear. Since the naive cosmographic expansion behaves much more moderately than the general expansion for e.g. observer 3 in model 3, we suspect that the main reason for the low convergence radius of the expansion in this region is the shear/high level of anisotropy, and in particular the gradients of the shear.
\newline\indent
We lastly note that the redshift along the light ray of observer 1 in model 4 is not monotonic. In such a situation, the redshift is not an appropriate expansion parameter. As discussed in \cite{recent_3}, one can in this situation consider the inverse expansion, i.e. expanding the redshift in terms of the angular diameter distance which is monotonic. We refrain from introducing such a scheme here since this is the only light ray we consider where the redshift turns out non-monotonic, and looking at figure \ref{Fig:ddl_model4} we do not see any evidence of striking effects from the non-monotonicity of the redshift.
\newline\newline\noindent
We do not attempt a formal computation of the radius of convergence here (this would require computing higher order expansion coefficients). However, we can follow the scheme used in appendix A of \cite{recent_1} and assess the radius of convergence by noting that near/somewhat after the redshift values close to the radius of convergence, the terms in the cosmographic expansion should be of similar order of magnitude; when the terms become of same order of magnitude, higher order terms are no longer small adjustments to the final estimate of the luminosity distance, but are rather modifying it at its leading order. From this, we estimate the order of the radius of convergence for the studied models as $z\sim D_L^{(i)}/D_L^{(i+1)}$, with the results presented in table \ref{table:convergence1-3}. The table gives the estimated orders of magnitude of the radius of convergence computed {\em only} based on the exact values of the expansion coefficients and not the naive counterparts. We provide the corresponding radius of convergence for the background FLRW model but note that this radius of convergence (which is 1) is only valid in a universe that is FLRW everywhere. The values in table \ref{table:convergence1-3} should be considered very crude upper limits which is perhaps best highlighted by the fact that we for several models and observers must include an interval because the fractions $D_L^{(i)}/D_L^{(i+1)}$ do not yield the same orders of magnitude for $i = 1$ and $i = 2$. Although it is the smallest number in each interval which best indicates the radius of convergence, we cannot assume that the estimate will not change drastically if we consider higher order coefficients. It is nonetheless somewhat surprising that the table indicates radii of convergence of the high value $z \sim  0.1$ for several models and observers despite figures \ref{Fig:ddl_model1}-\ref{Fig:ddl_model4} clearly demonstrating that the radius of convergence has been surpassed at $z = 0.1$ in nearly all cases studied. It is additionally notable that although the estimates in general seem to overestimate the true radius of convergence, there is somewhat agreement between the lowest estimates and the clearest and earliest actual divergence of the cosmographic expansions. This is in particular true for the light ray of observer 3 in model 4 which has the lowest radius of convergence estimate and which also shows the clearest and earliest divergence of the cosmographic expansion (see figure \ref{Fig:ddl_model4}). We nonetheless all-in-all conclude that the estimates of the radii of convergence in table \ref{table:convergence1-3}  must be considered very crude and generally a bit too optimistic.
\newline\newline\noindent
{\bf\underline{Models 5 and 6: Living on the edge (of The Local Void)}}\newline
As seen in figure \ref{Fig:ray_rho}, observer 3 was in models 3 and 4 placed at the edge of the void, in a slight overdensity. These are also the observers for whom the general cosmographic expansion has an especially short radius of convergence (see figure \ref{Fig:ddl_model3}). Since our very local cosmic environment places us at the edge of The Local Void, it is interesting to examine the convergence of the general cosmographic expansion with a few more examples of observers placed near the edge of a void. Specifically, we consider present-time observers placed in the overdensities of models 5 and 6, with observers always having $k^r<0$ but otherwise random lines of sight. We specifically choose the observer positions $r = 25, 27, 30$, (observer 4, 5 and 6, respectively) corresponding to placing the observer on the  ``void-side'' of the overdensity, near the top of the overdensity and ``behind'' the overdensity, close to the FLRW background. Since models 5 and 6 represent structures with radius of only $36$Mpc, we only trace the light rays up to $z = 0.02$.
\newline\indent
\begin{table}[!htb]
	\centering
	\begin{tabular}{c c c c c c c c}
		\hline\hline
		Model &  $\mathcal{H_O}$ (km/s/Mpc) &  $\mathcal{Q_O}$ & $\mathcal{Q}_{\rm naive}$ & $\mathcal{R_O}$ & $\mathcal{R}_{\rm naive} $ & $\mathcal{J_O}$& $\mathcal{J}_{\rm naive}$\\
		\hline
		\Tstrut
		Model 5, obs4 & 61.9  & 394 & -0.494 & 394  & -0.680 & $-370\cdot 10^{3}$   & 1.71    \\
		Model 5, obs5 & 57.2  & 239 & -0.403  & 239  & -1.27 & $185\cdot 10^4$ & 2.23       \\
		Model 5, obs6 & 62.0  & -164 & -0.524 & -164 & -0.647 & $291\cdot 10^2$  & 1.66   \\
		Model 6, obs4 & 47.4  & 412$\cdot 10$ & 0.342  & 411$\cdot 10$ & -3.44  & $-397\cdot 10^5$  & 3.09        \\
		Model 6, obs5 & 21.4  & 151$\cdot 10$ & 36.2  & $150\cdot 10$  & -69 & $182\cdot 10^7$   & -394    \\
		Model 6, obs6 & 51.8  & -163$\cdot 10$ & -0.197 & -163$\cdot 10$  & -2.23 & -116$\cdot 10^5$   & -0.197    \\
		FLRW limit & 70.0 & -0.55  & -& 0 & - & 1 & - \\
		\hline
	\end{tabular}
	\caption{Cosmographic expansion parameters for models 5 and 6 for the three observers together with the FLRW limit of the considered LTB models.}
	\label{table:Models5-6}
\end{table}

\begin{table}[!htb]
	\centering
	\begin{tabular}{c c }
		\hline\hline
		Model &  radius of convergence ($z$) \\ 
		\hline
		\Tstrut
		Model 5, obs4 &   $10^{-3}$ \\
		Model 5, obs5 &     $10^{-4}-10^{-2}$\\
		Model 5, obs6 &  $10^{-2}$\\
		Model 6, obs4 &  $10^{-4}$\\
		Model6, obs5 &  $10^{-5}-10^{-4}$\\
		Model 6, obs6 &  $10^{-4}-10^{-3}$\\
		FLRW limit &  1\\
		\hline
	\end{tabular}
	\caption{Estimated order of magnitude of the radius of convergence for models 5 and 6 for the three observers together with the FLRW background.}
	\label{table:convergence5-6}
\end{table}

\begin{figure}[]
	\centering
	\includegraphics[width=0.45\linewidth]{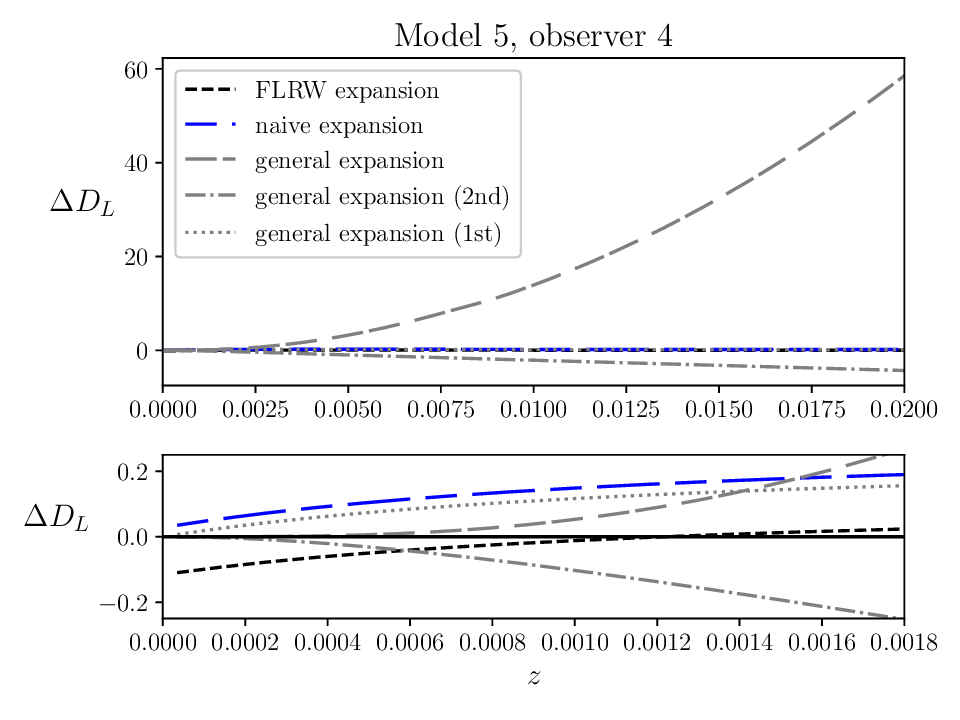}
	\includegraphics[width=0.45\linewidth]{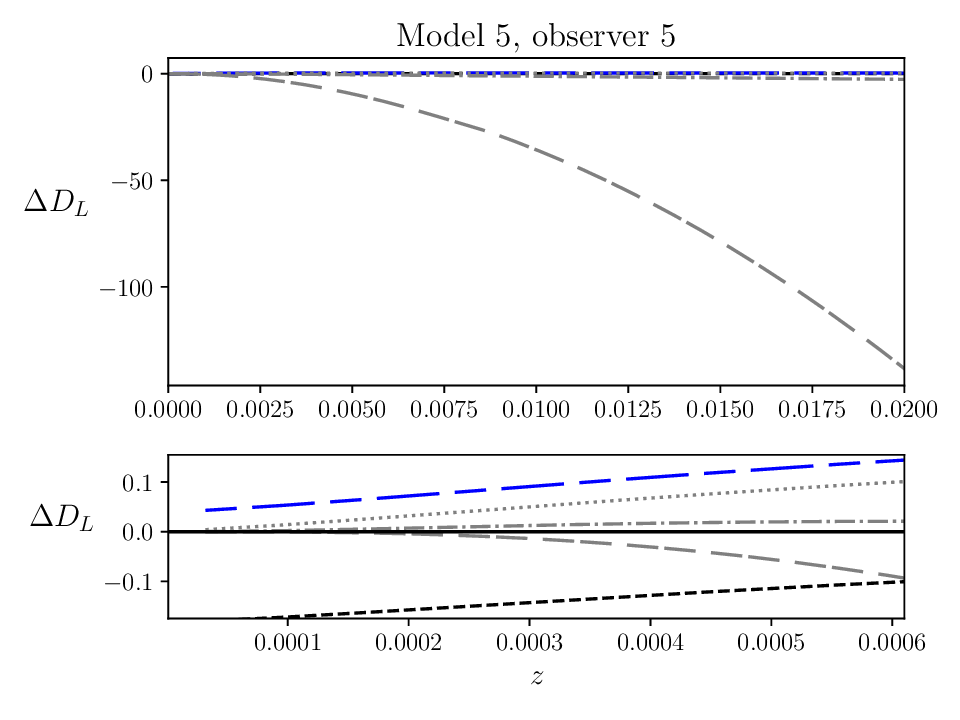}
	\includegraphics[width=0.45\linewidth]{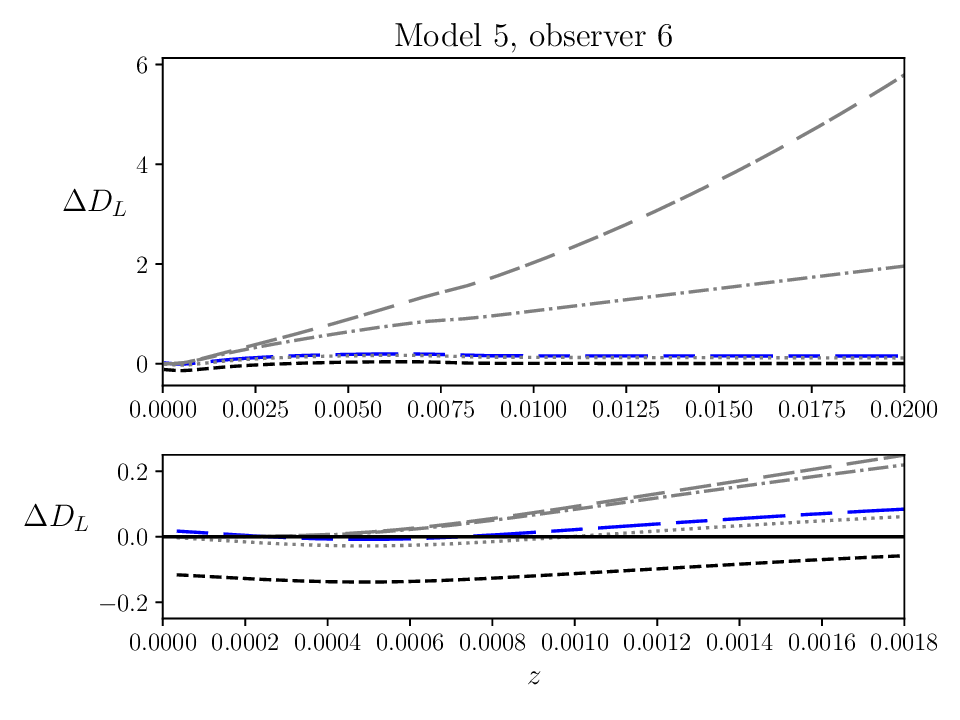}
	\caption{Cosmographic expansions shown as relative deviation from the exact luminosity distance, $\Delta D_L :=({\rm cosmographic\,\,\, expansion}-D_L)/D_L$, along fiducial lines of sight for three present-time observers in model 5. The naive and FLRW cosmographic expansions are only shown at third order. The general expansion is shown at first, second and third order. An inset with a black zero-line to show the low-z behavior of the expansions.}
	\label{Fig:ddl_model5}
\end{figure}

\begin{figure}[]
	\centering
	\includegraphics[width=0.45\linewidth]{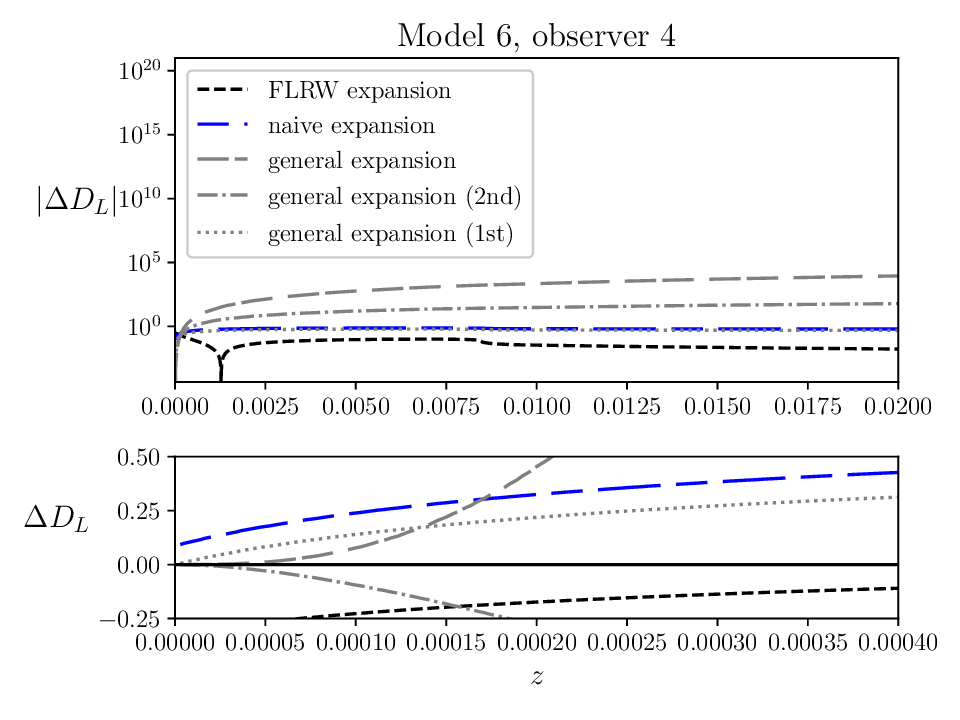}
	\includegraphics[width=0.45\linewidth]{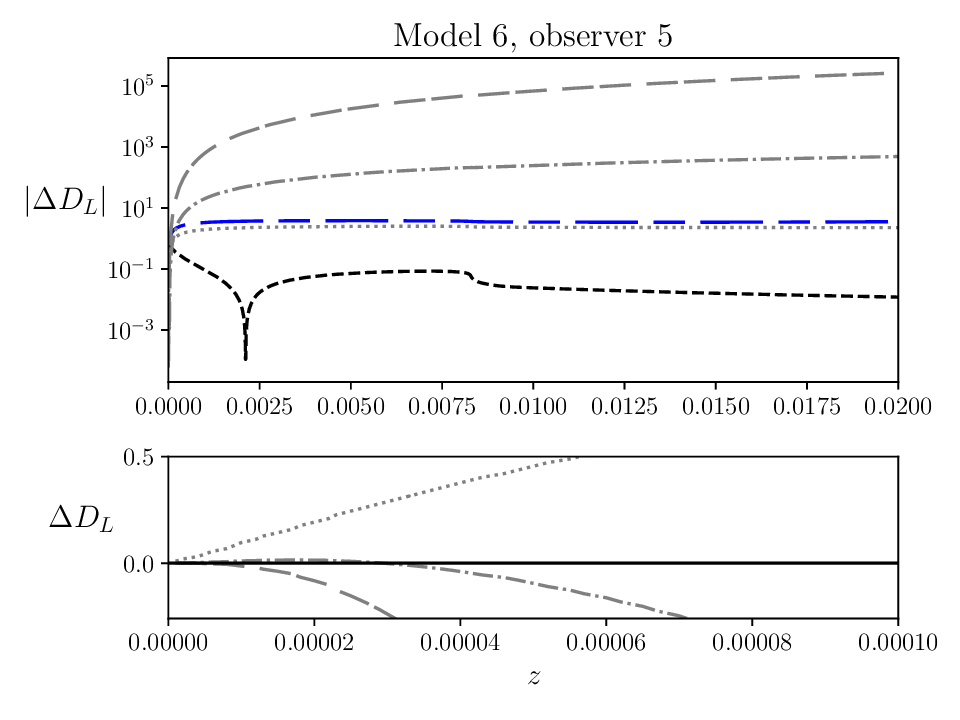}
	\includegraphics[width=0.45\linewidth]{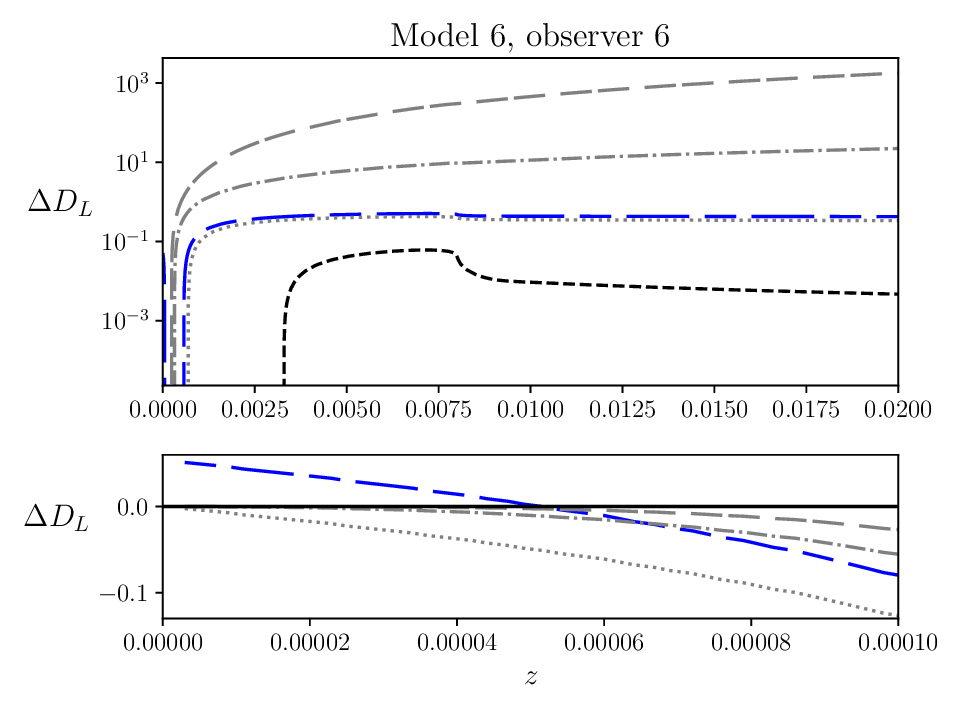}
	\caption{Cosmographic expansions shown as relative deviation from the exact luminosity distance, $\Delta D_L :=({\rm cosmographic\,\,\, expansion}-D_L)/D_L$, along fiducial lines of sight for three present-time observers in model 6. The naive and FLRW cosmographic expansions are only shown at third order. The general expansion is shown at first, second and third order. An inset with a black zero-line to show the low-z behavior of the expansions.}
	\label{Fig:ddl_model6}
\end{figure}
The expansion coefficients for each observer in the two models are shown in table \ref{table:Models5-6}. We note that for the chosen observers, the effective Hubble constant $\mathcal{H_O}$ can deviate from the $\Lambda$CDM background value by as much as $\sim 70\%$ while the effective deceleration parameter deviates up to several hundreds of percent from the background value. Even the naive value of $\mathcal{Q_O}$ deviates by this much for one observer in one model (observer 5 in model 6). We also notice that the naive extrapolation of $\mathcal{Q_O}$ now also becomes positive for some observers but even in these cases, the general value of $\mathcal{Q_O}$ and its naive extrapolation are very different. This is also the case for $\mathcal{J_O}$ and its naive extrapolation which e.g. do not always have the same sign. We also note that we now find that $\mathcal{Q_O}$ and $\mathcal{R_O}$ are almost identical for all observers in both models. This is contrary to the results for models 1-4 where this was only the case for a single observer in a single model. We can understand this by noticing that $|\mathcal{Q_O}|$ is now so large that it, apparently, dominates the terms in $\mathcal{R_O}$.
\newline\indent
\begin{figure}[]
	\centering
	\includegraphics[width=0.45\linewidth]{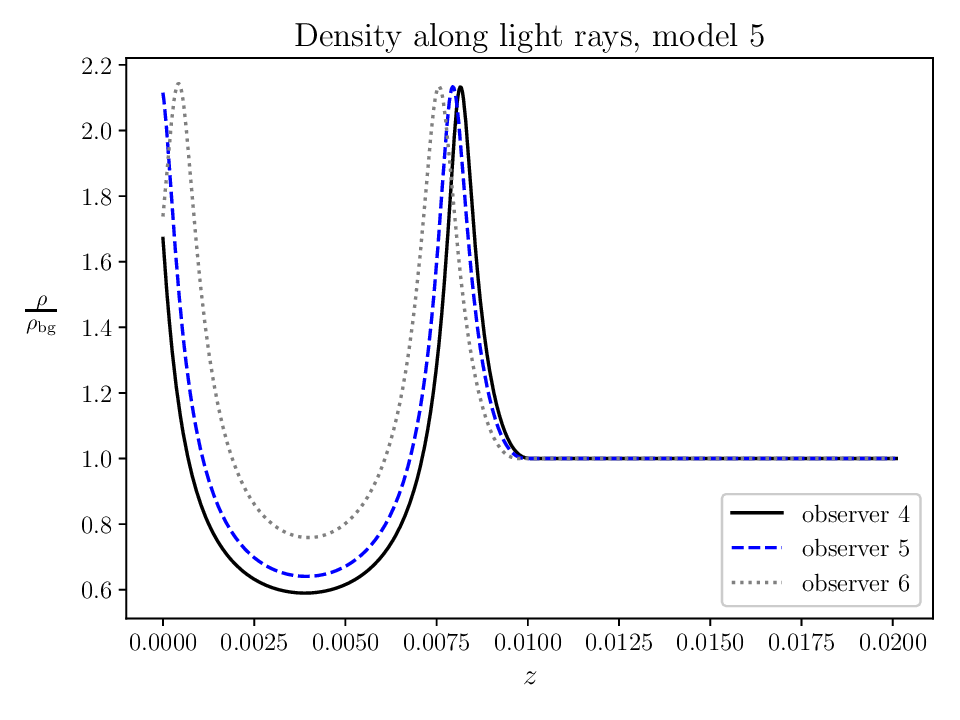}
	\includegraphics[width=0.45\linewidth]{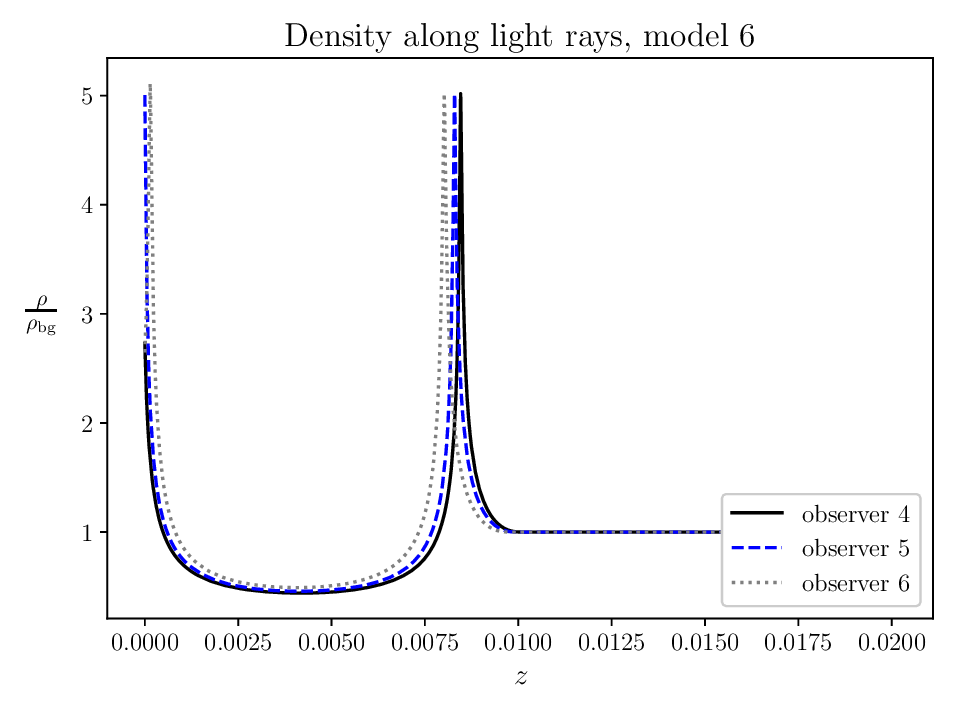}
	\caption{Density along the light rays studied for model 5 and 6.}
	\label{Fig:ray_rho5}
\end{figure}
The cosmographic expansions along each considered light ray are shown in figures \ref{Fig:ddl_model5} and \ref{Fig:ddl_model6}. We again clearly see that the third order general expansion diverges already at very low redshift and thus that the radius of convergence is very small for these models. The density profiles along the light rays are shown in figure \ref{Fig:ray_rho5} from which it is clear that the redshift is monotonic along all rays studied in models 5 and 6 (this can also be seen in figures \ref{Fig:ddl_model5} and \ref{Fig:ddl_model6}).
\newline\newline
As for the earlier models, we also here compute the crude estimates of upper limits of the radii of convergence for the cosmographic expansions. These are shown in table \ref{table:convergence5-6}. We again note that the estimated orders of magnitude are given as intervals because the estimates depend on whether we consider the first and second order coefficients or the second and third order coefficients. It is the lowest value in the interval which should be considered the best estimate of the radius of convergence. The estimates in table \ref{table:convergence5-6} seem to be in fair agreement with the plots in figures \ref{Fig:ddl_model5} and \ref{Fig:ddl_model6}, but we nonetheless caution against taking these numbers at face values since we clearly found for models 1-4 that the estimates are simply too crude to be considered particularly useful.
\newline\newline
{\bf\underline{Models 5-6: 100 lines of sight for observer 4}}\newline
So far, we have only considered three random lines of sight per observer and model. We therefore risk that the results presented above by chance significantly under or over represent the effects of the local environment on the convergence of the cosmographic expansions. To assess if this is the case, we here consider 100 random lines of sight (now also permitting $k^r>0$) for observer 4 in models 5 and 6, using the same random sequence of lines of sight for both models. We choose these specific models and this observer since the corresponding local environment is among the most anisotropic of those studied.
\newline\indent

\begin{figure}[]
	\centering
	\includegraphics[width=0.45\linewidth]{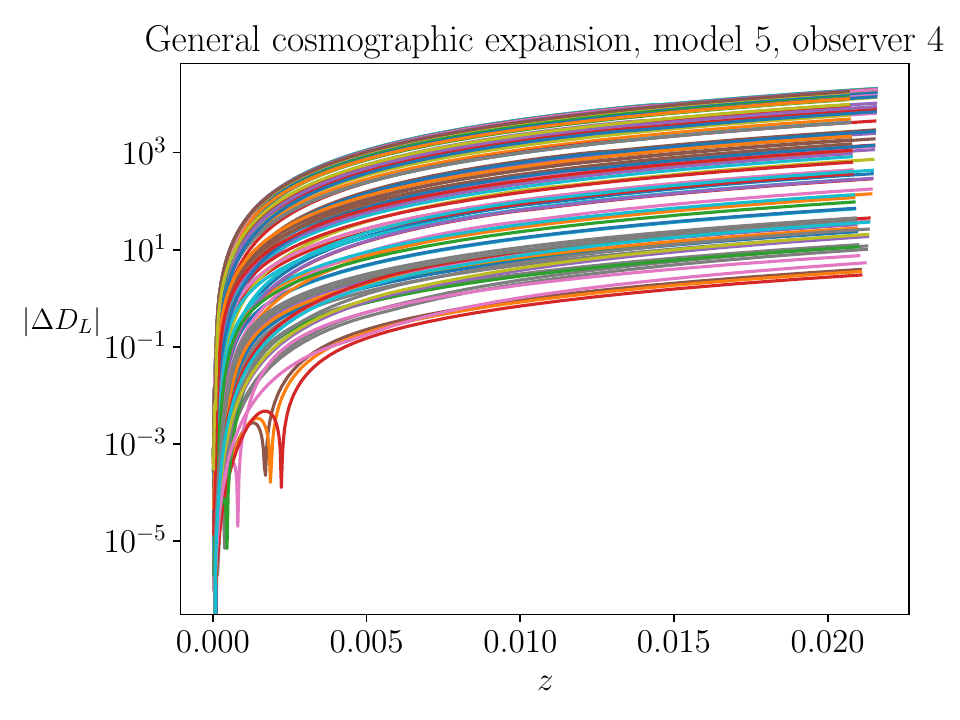}
	\includegraphics[width=0.45\linewidth]{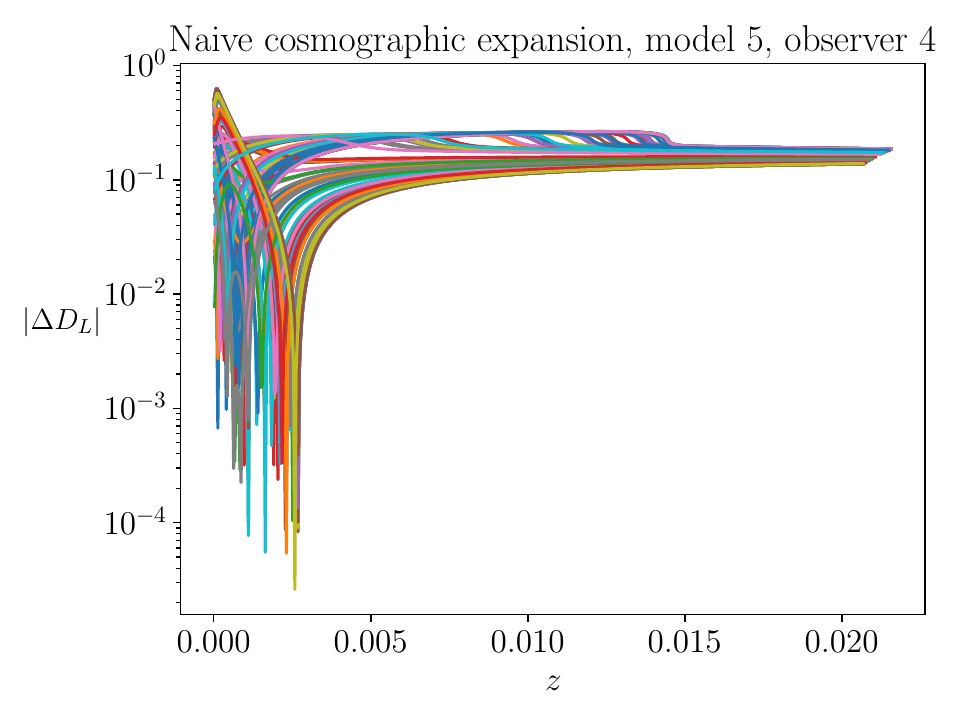}
	\includegraphics[width=0.45\linewidth]{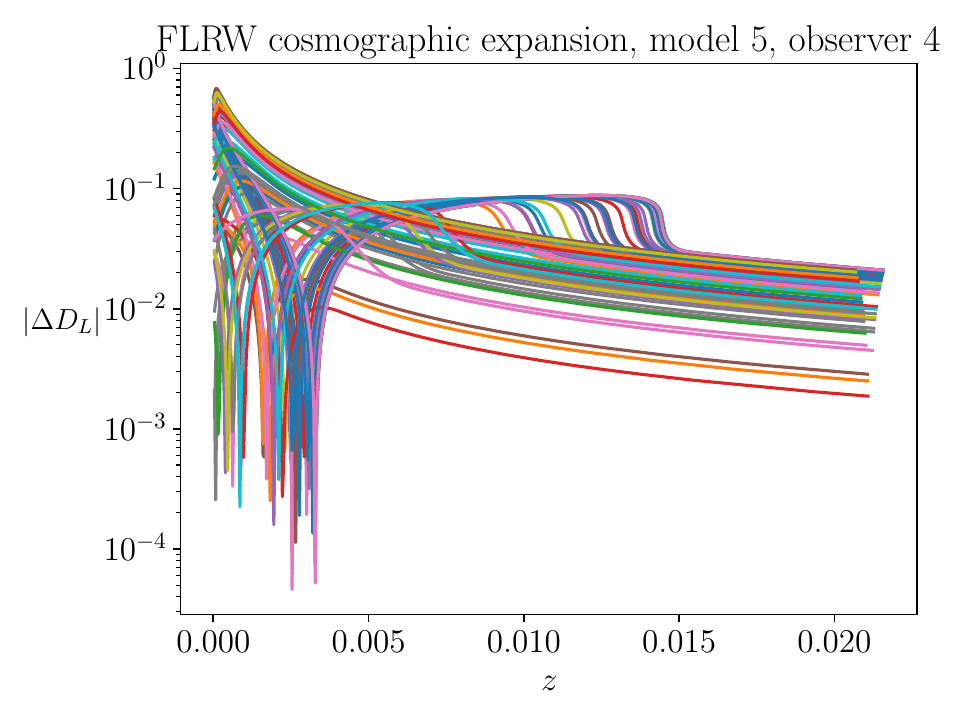}
	\caption{Cosmographic expansions shown as relative deviations from the exact luminosity distance, $\Delta D_L :=({\rm cosmographic\,\,\, expansion}-D_L)/D_L$, along 100 fiducial lines of sight for observer 4 in model 5. All cosmographic expansions are to third order.}
	\label{Fig:50_model5}
\end{figure}
\begin{figure}[]
	\centering
	\includegraphics[width=0.45\linewidth]{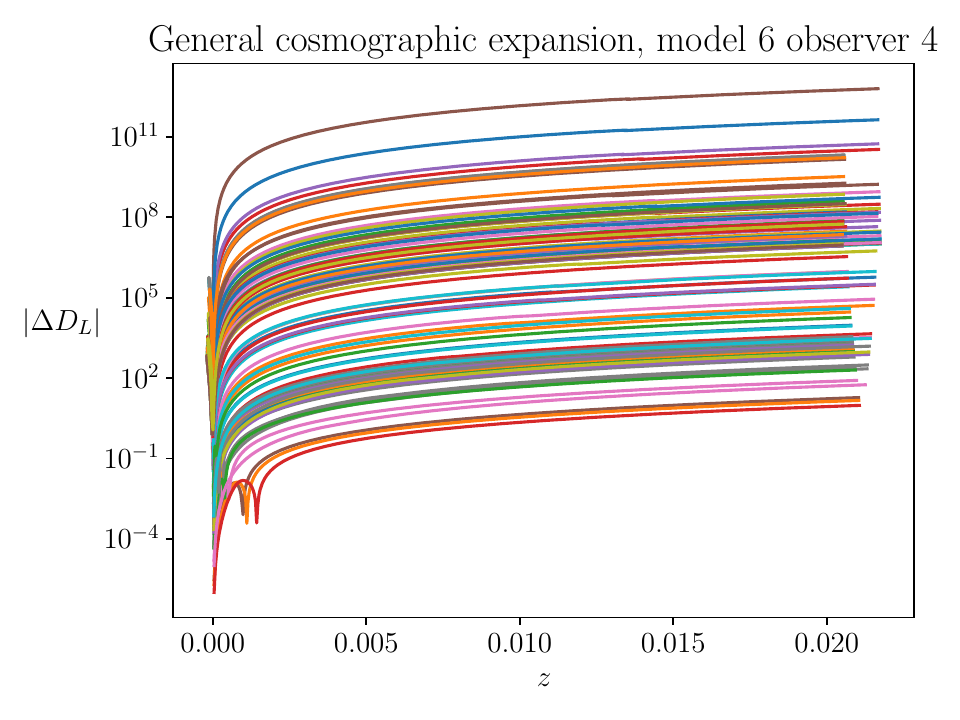}
	\includegraphics[width=0.45\linewidth]{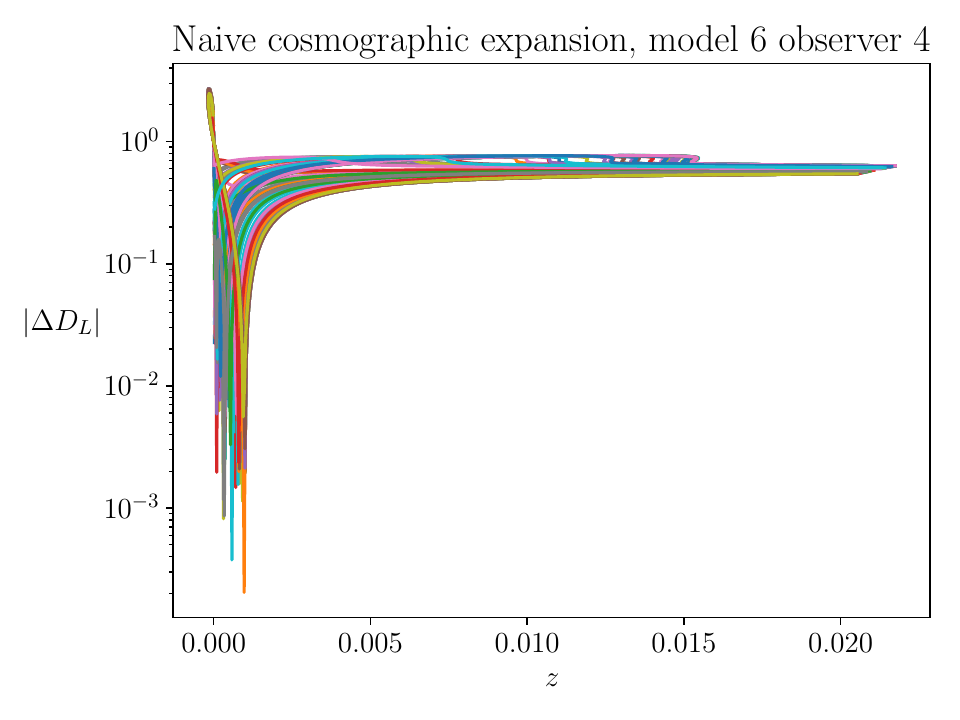}
	\includegraphics[width=0.45\linewidth]{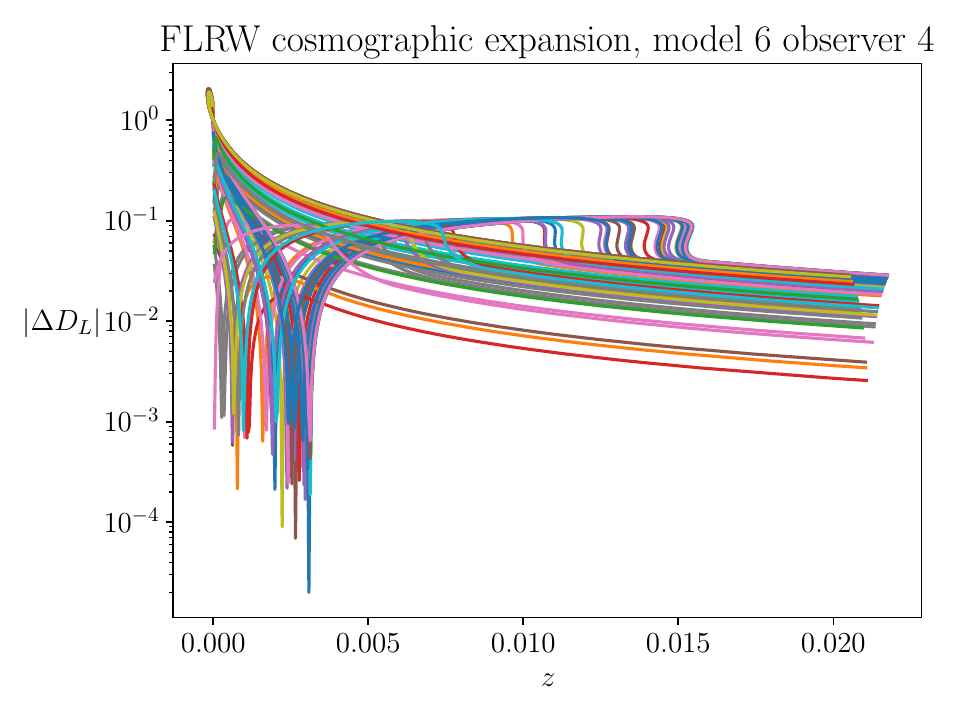}
	\caption{Cosmographic expansions shown as relative deviations from the exact luminosity distance, $\Delta D_L :=({\rm cosmographic\,\,\, expansion}-D_L)/D_L$, along 100 fiducial lines of sight for observer 4 in model 6. All cosmographic expansions are to third order.}
	\label{Fig:50_model6}
\end{figure}

\begin{figure}[]
	\centering
	\includegraphics[width=0.45\linewidth]{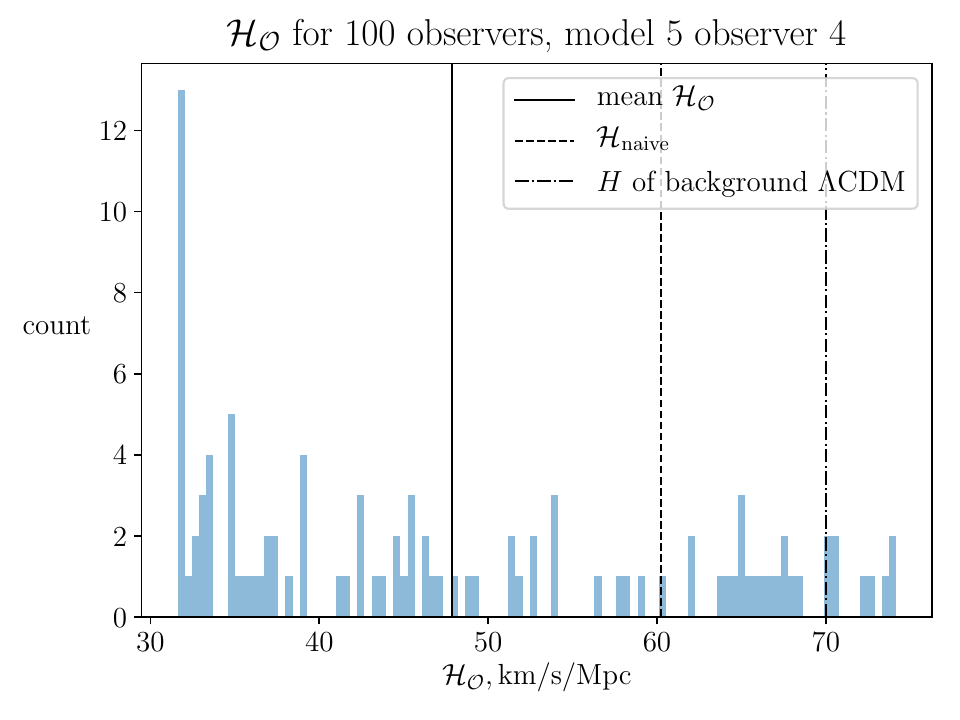}
	\includegraphics[width=0.45\linewidth]{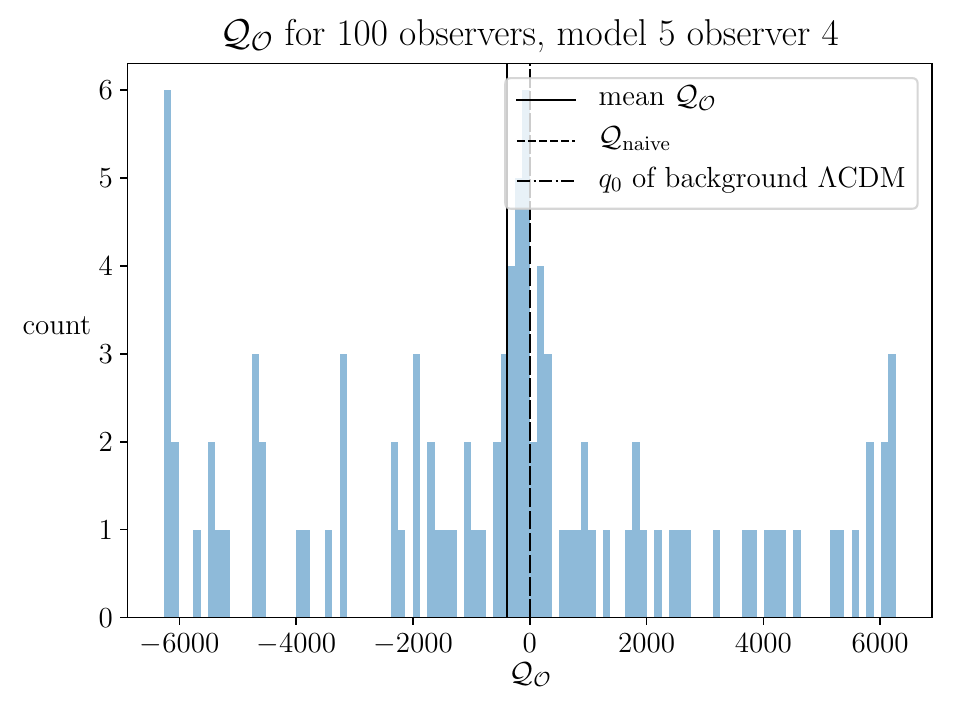}\\
	\includegraphics[width=0.45\linewidth]{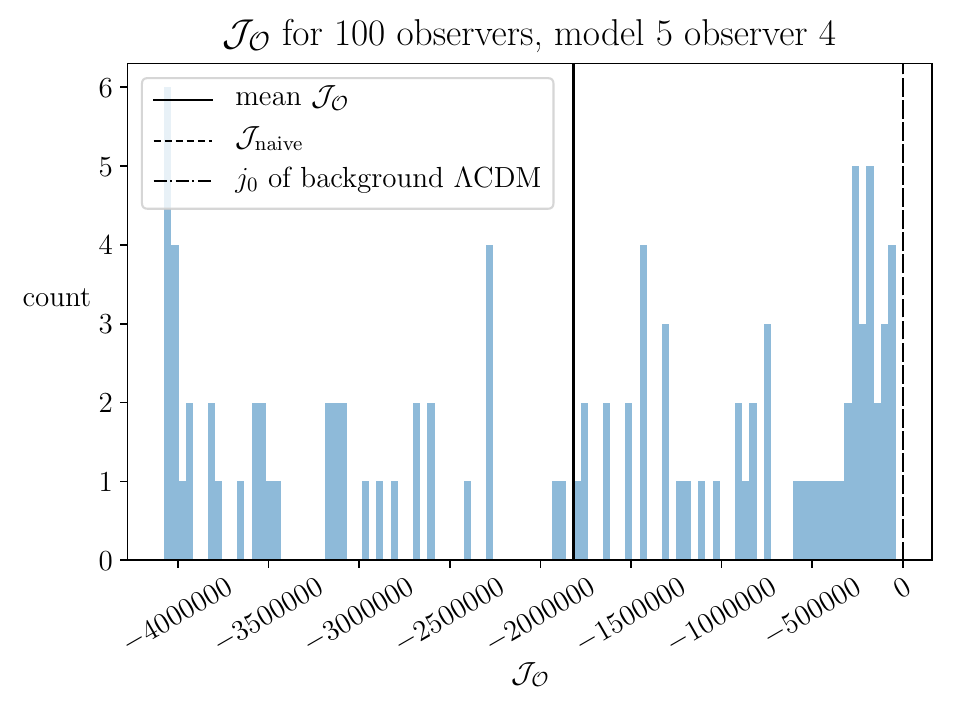}
	\includegraphics[width=0.45\linewidth]{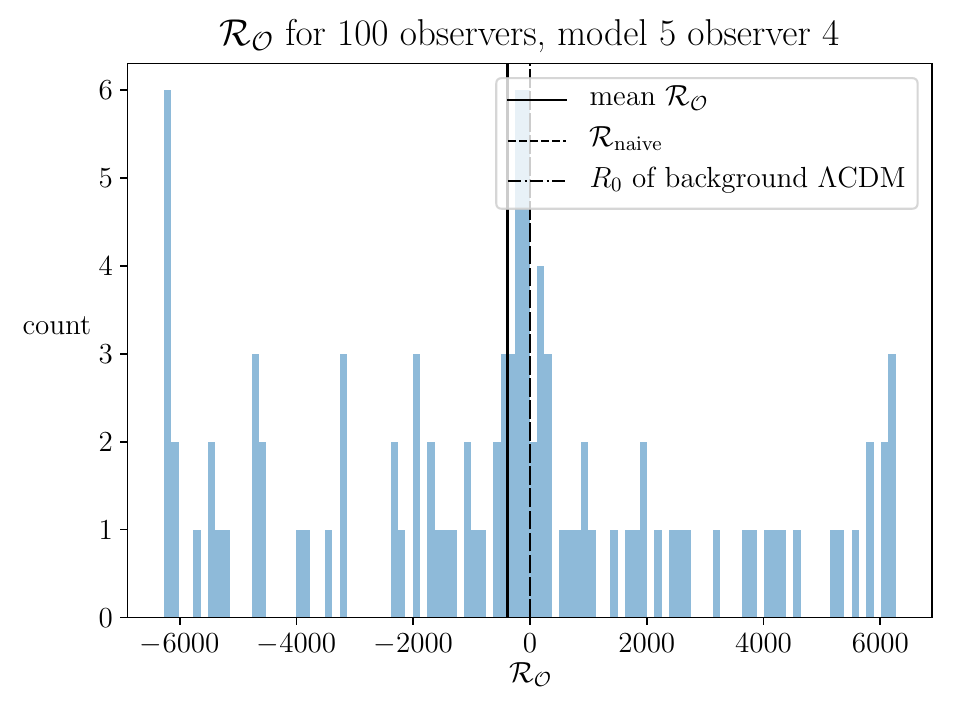}
	\caption{Distributions of $\mathcal{H_O} (in km/s/Mpc), \mathcal{Q_O}, \mathcal{J_O}$ and $\mathcal{R_O}$ along 100 lines of sight for observer 4 in model 5. The histograms also show the mean values and naive and FLRW limits which are independent of the lines of sight. Except for $\mathcal{H_O}$, the values of the FLRW limit and its naive extrapolation cannot be distinguished in the histograms due to the large spread of the general value.}
	\label{Fig:hist_modddel5}
\end{figure}

The deviations between the third order cosmographic expansions and exact redshift-distance relation along the 100 lines of sight for observer 4 in each model are shown in figures \ref{Fig:50_model5} and \ref{Fig:50_model6}. The figures include both the general expansion as well as the FLRW expansion and its naive isotropic limit. For the general cosmographic expansion we find that the deviation from the exact luminosity distance varies significantly between the different lines of sight. Although there are also variations between individual lines of sight for the FLRW expansion and it naive extrapolation, the fluctuations are modest compared to the general cosmographic expansion. As for the single light rays studied above, we here see that the FLRW expansion becomes better at higher redshift. We remind the reader that this is expected for LTB models where the redshift-distance relation reduces very closely to that of the background once the light ray enters the background. There is therefore a priori no reason to expect that similar behavior would be seen in a more general spacetime without an explicit ``background''.
\newline\indent
Figures \ref{Fig:50_model5} and \ref{Fig:50_model6} reveal that there do exist lines of sight along which the deviation between the general cosmographic expansion and the exact redshift-distance relation is fairly small (of order less than or around 1). However, these lines of sight appear to be special and the deviation is several orders of magnitude above 1 for most lines of sight already at very low redshifts.
\newline\newline
Averaging over sources to some extent defeats the purpose of general cosmographic expansions where the hope is to measure dynamical degrees of freedom in our local cosmic neighborhood \cite{general_1}. Nonetheless, a typical cosmological observational survey naturally covers more than a single line of sight and hence it will typically be possible to consider source averages in relation to cosmographic expansions. As the references in the introduction illustrate, the actual number of sources and hence lines of sight depends heavily on the type(s) of source(s) considered. If e.g. supernovae samples are considered, the number of sources can be in the hundreds to thousands, but if instead BAO data is considered, the number of sources will be significantly less and only of the order 10. We have here considered 100 lines of sight since we expect this will be enough to qualitatively understand the effects of averaging over several lines of sight. In figure \ref{Fig:hist_modddel5} we show the distributions of $\mathcal{H_O}, \mathcal{Q_O}, \mathcal{R_O}$ and $\mathcal{J_O}$ along the 100 lines of sight of observer 4 in model 5. The figure also shows their mean values, the FLRW limit and its naive extrapolation. The figure shows that there is significant spread in the general values and that the mean does not reduce to neither the FLRW or naive limits. While it is expected that the general expressions would not average out to their FLRW limits, it is perhaps more surprising that that they do not even average out to their naive counterparts. However, except for $\mathcal{H_O}$, this is actually expected as it was shown analytically in \cite{general_1} that  the monopole limit of $\mathcal{Q_O}$ does not reduce to the naive extrapolation of the FLRW deceleration parameter. On the other hand, the effective Hubble parameter is at least naively expected to reduce to the naive extrapolation of the FLRW limit upon averaging. However, since the spread in the $\mathcal{H_O}$ values is so large, we would need significantly more lines of sight to obtain a true estimate of the mean of $\mathcal{H_O}$ over the sky. This is even more so for $\mathcal{J_O}$ and $\mathcal{R_O}$ which fluctuate over several orders of magnitude, indicating that we would need several orders of magnitude more lines of sight to obtain a stable mean.
\newline\indent

\begin{figure}[]
	\centering
	\includegraphics[width=0.45\linewidth]{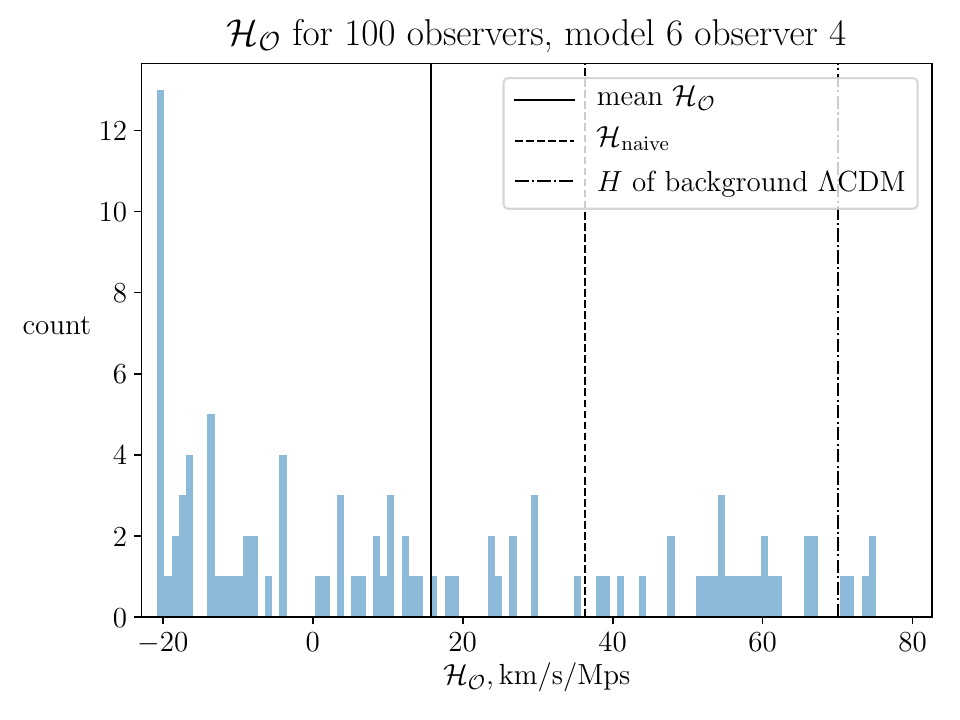}
	\includegraphics[width=0.45\linewidth]{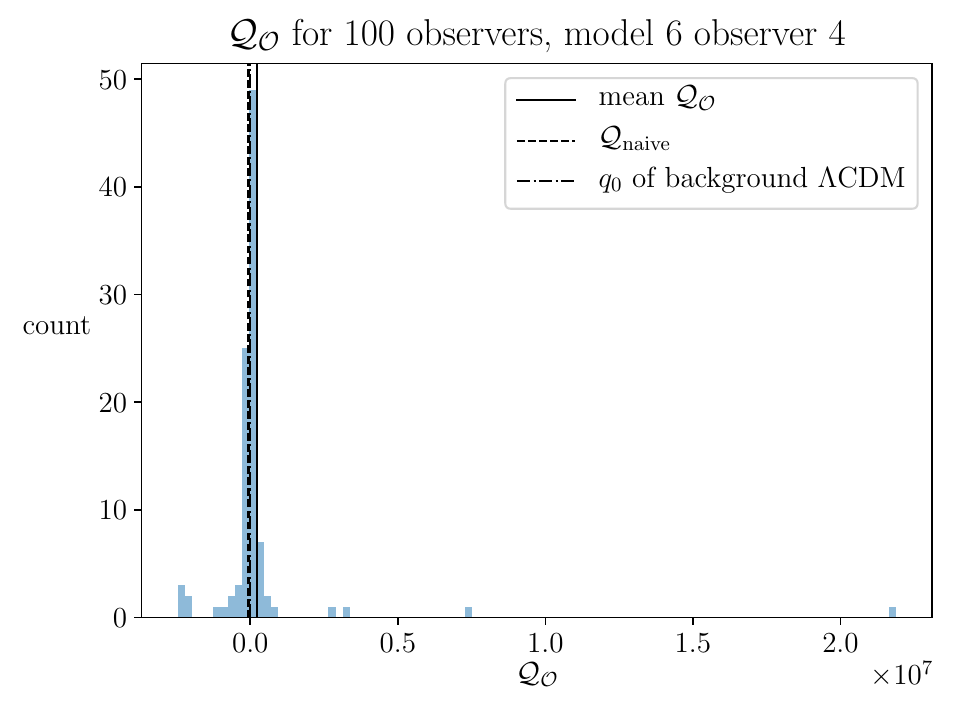}
	\caption{Distributions of $\mathcal{H_O}$ (in km/s/Mpc) and $\mathcal{Q_O}$ along 100 lines of sight for observer 4 in model 6. The histograms also show the mean values and naive and FLRW limits which are independent of the lines of sight. }
	\label{Fig:hist_modddel6}
\end{figure}

The variations in $\mathcal{H_O}$ and $\mathcal{Q_O}$ in model 6 are shown in figure \ref{Fig:hist_modddel6}. These results are consistent with those for model 5, but the fluctuations are now larger and we find the variations in $\mathcal{J_O}$ and $\mathcal{R_O}$ to be over so many order of magnitude that it becomes difficult to assess the information from a simple plot. We therefore do not show the corresponding histograms. We also note that in the case of model 6, $\mathcal{H_O}$ can take both signs, which we did not find among the light rays in model 5 where $\mathcal{H_O}$ is positive for all studied light rays.
\newline\newline
From studying the 100 random lines of sight in models 5 and 6 we overall conclude that although there certainly are special lines of sight in these models, our earlier results based on 3 lines of sight per observer are robust in the sense that the general cosmographic expansion can have a very small radius of convergence for some inhomogeneous models such as LTB models.

\subsection{Comparison with earlier work}
In this subsection we compare our results explicitly to recent work by others considering cosmographic expansions that take inhomogeneities of the cosmological model into account.
\newline\newline
In \cite{recent_1} the authors computed the luminosity distance in cosmological simulations obtained with numerical relativity. The authors found around 2\% and 120\% maximum deviations between the background versus effective observational Hubble ($\mathcal{H_O}$) and deceleration ($\mathcal{Q_O}$) parameters, respectively. Although this is significantly less than what we find here, the simulations considered in \cite{recent_1} also only contained inhomogeneities of order 0.05 \% which is significantly less than what we focus on here.
\newline\indent
The convergence of the general cosmographic expansion was estimated in appendix A of \cite{recent_1}, based on the method used in \cite{better_1} to consider the convergence of the FLRW cosmographic expansion. It was concluded that for the considered models, where structures were smoothed below the scale of $100$Mpc/$h$, the general cosmographic expansion should be valid only up to a redshift of $\sim 0.02- 0.03$. This is at least an order of magnitude above the radius of convergence for most observers and models studied here, as indicated by the plotted luminosity distance and cosmographic expansions along the considered light rays (although it is pretty close to the order of magnitude we expect for observer 1 in model 1 where the inset indicates that the third order expansion is significantly better than the second order expansion up to around this redshift). The estimate of $z\sim 0.02-0.03$, however, fits within the overall interval of radii of convergence we estimate for the considered models presented in tables \ref{table:convergence1-3} and \ref{table:convergence5-6}. Since we have significantly larger density contrasts than considered in \cite{recent_1} (also when taking into account that we consider a $\Lambda$CDM rather than an EdS background which is considered in \cite{recent_1}), we consider it reassuring that even for the relatively large density contrast of model 1 (compared to those studied in \cite{recent_1}), the estimates of \cite{recent_1} are in reasonable agreement with our results. We also note that the estimates of appendix A in \cite{recent_1} were verified numerically in \cite{Hayley}. However, it was noted in \cite{Hayley} that when smaller scales are included, the cosmographic expansion is much less successful at reproducing the true redshift-distance relation and it was suggested that a method to overcome this when applying the general cosmographic expansion to the real universe could be to introduce smoothing scales. Our results support this suggestion. In particular, our results indicate that the main obstacle for the convergence of the general cosmographic expansion is the large derivatives of the anisotropic quantities such as the shear. By introducing smoothing scales, these derivatives would become smaller and hence the convergence radius of the expansion would increase. This has a price though, since the smoothing reduces the amount of information that we can extract from the general cosmographic expansion and in addition, it is not clear how this smoothing should be performed in practice with real data.
\newline\indent
In \cite{general_3}, the authors consider the cosmographic expansion of a parameter $\eta$ related to the luminosity distance. The authors find that the convergence of those cosmographic expansions can become poor due to non-linearities. This is similar to what is found here, except that we find that it does not seem to be the nonlinearity/large density contrast itself but rather the variation of the anisotropic features of the spacetime that lead to the breakdown of the cosmographic expansion. In \cite{general_3}, the authors also note that higher order approximations in general do better than the lower order approximations. In that case, the lack of convergence of the expansion could simply be because higher order terms are necessary for the expansion to make a good approximation. This is not the case for the results found here where the higher order approximation clearly diverges also much faster than the lower order expansions, which is a clear sign that the radius of convergence has been exceeded.
\newline\indent
Lastly, we remark on the work presented in \cite{2nd}. There, the authors consider a perturbed universe using second order perturbation theory and compute the corresponding redshift-distance relation. The authors afterwards fit these ``observed'' redshift-distance relations to FLRW cosmographic expansions, finding large discrepancies between the Hubble and deceleration parameters of the background model versus those inferred from the cosmographic fit for sufficiently large local structures. A somewhat similar fit of FLRW cosmographic expansions to the exact redshift-distance relations in various Szekeres models were performed in \cite{selv}, where percent level deviations between inferred and background parameter values were identified. These studies are somewhat different than the study performed here, but we can nonetheless say that they overall agree with the results found here, where the effective values of the Hubble and deceleration parameters listed in tables \ref{table:Models1-3} and \ref{table:Models5-6} deviate (significantly) from the background values.

\section{Summary and conclusions}\label{sec:conclusion}
We constructed six different LTB models with $\Lambda$CDM backgrounds. For each  model, we considered light rays for three observers, computing the exact redshift-distance relation along the light rays as well as cosmographic approximations of this relation. We considered and compared three types of cosmographic expansions, namely the standard FLRW expansion introduced in \cite{jerk_1, jerk_2}, the general expansion of \cite{general_1} and the naive, isotropic extrapolation of the FLRW expansion. We generally find that the local values of the effective Hubble, deceleration, jerk and curvature parameters can vary by several orders of magnitude compared to the background. The main contribution to the large deviations is local anisotropy and in particular the gradient of the shear. We conclude this by noting that the naive extrapolations of the Hubble, deceleration, jerk and curvature parameters are generally much closer to the background values than the general expressions. We also find that the radius of convergence of the general cosmographic expansion is small in all the studied cases. Although we do not compute the expansion coefficients to high enough order to estimate the radius of convergence, the divergence of the cosmographic expansions and the higher order expansions becoming worse than the lower order expansions reveals that the radius of convergence has been exceeded. This is in line with tentative results obtained by others using analytical estimates, perturbation theory and simulations. We do include crude estimates of upper limits of the radius of convergence but conclude that the values in many cases seem much too large to be useful when comparing the the actual convergence of the cosmographic expansions.
\newline\indent
For most of the studied observers and models we only consider three random lines of sight per observer. To ensure that these three lines of sight (per observer) do not severely under- or over represent the effects of anisotropy, we extended the study with 100 random lines of sight for one of the observers in models 5 and 6. We found that the different lines of sight resulted in significant variations of the effective Hubble, deceleration, jerk and curvature parameters, for the latter three of several orders of magnitude. Although we did find some lines of sight where the general cosmographic expansion diverged only modestly (below order 1) from the exact redshift-distance relation, these seem to be special rays. The general conclusion based on the 100 lines of sight is still that the general cosmographic expansion has a very small radius of convergence.
\newline\indent
We included a computation of average effective Hubble, deceleration, jerk and curvature parameters along the 100 lines of sight each for the one observer in models 5 and 6. We naively expect that at least $\mathcal{H_O}$ would average to the naive extrapolation of the FLRW value, but do not find this for our sample. We do not find that the general deceleration, jerk and curvature parameters reduce to their naive counter parts either, but in these cases this is as expected as it was shown in \cite{general_1} that the monopole limits of the general expressions do not reduce to their naive FLRW extrapolations. Since the spread is very large (several order of magnitude for all but $\mathcal{H_O}$), we expect that the means we compute can change significantly if more light rays are considered and that we would need several orders of magnitude more light rays to obtain faithful estimates of the means.
\newline\indent
The low radius of convergence and the general cosmographic expansion is a possible obstacles for using general cosmographic expansions with real data. At the very least, a necessary step before using the expansion with real data should be to estimate the convergence of the expansion with realistic simulations of our local cosmic neighbourhood as also suggested in \cite{Hayley}. This would also be important for assessing the validity of the suggested framework in the preprint \cite{chris} where the authors claim to show that their general cosmographic expansion is suitable up to $z\lesssim 0.1$. This assessment may be too optimistic as such high redshift may be much beyond the radius of convergence of general cosmographic expansions.

\clearpage
\section{Acknowledgments}
We thank Asta Heinesen for correspondence and encouragement to look into the convergence of cosmographic expansions.
\newline\newline
This project was funded by VILLUM FONDEN, grant VIL53032.
\newline\newline
{\bf Author contribution statement:} The numerical results were obtained by modifying preexisting code written by SMK. Both authors contributed to the modifications and debugging of the code written for the project as well as to the analytical derivations. The writing of the manuscript was led by SMK with contributions from ABM.
%\newpage

\end{document}